\documentclass[useAMS,usenatbib]{mn2e}

\usepackage{natbib, aas_macros}
\usepackage{version} 
\usepackage{ulem}
\usepackage{graphicx}
\usepackage{wrapfig}
\graphicspath{{./figs/}}
\usepackage{color}
\usepackage{amsmath}
\usepackage{amssymb}

\newcommand{\Msun}{{\rm  M_{\odot}}}

\newcommand{\Zsun}{Z_{\odot}}

\newcommand{\Lsun}{L_{\odot}}

\newcommand{\lya}{\rm {Ly{\alpha}}}

\newcommand{\Mh}{M_{\rm h}}
\newcommand{\Msunyr}{\rm M_{\odot}~ yr^{-1}}

\newcommand{\cc}{{\rm cm^{-3}}}

\newcommand{\un}[2]{#1_{\rm #2}}

\newcommand{\III}{I\hspace{-.1em}I\hspace{-.1em}I}
\newcommand{\II}{I\hspace{-.1em}I}
\newcommand{\cii}{{\sc [C\,i\hspace{-.1em}i]} }
\newcommand{\oiii}{{\sc [O\,i\hspace{-.1em}i\hspace{-.1em}i]} }

\title[Metal emission lines from the first galaxies] 
{Starbursting \oiii emitters and quiescent \cii emitters in the reionization era
}
\author[Arata et al.]
{Shohei Arata$^{1}$\thanks{E-mail: arata@astro-osaka.jp},
Hidenobu Yajima$^{2}$,
Kentaro Nagamine$^{1,3,4}$,
Makito Abe$^{2}$\and
and Sadegh Khochfar$^{5}$
\\
$^{1}$ Department of Earth and Space Science, Graduate School of Science, Osaka University, Toyonaka, Osaka 560-0043, Japan\\
$^{2}$ Center for Computational Sciences University of Tsukuba, Ibaraki 305-8577, Japan\\
$^{3}$ Department of Physics \& Astronomy, University of Nevada, Las Vegas, 4505 S. Maryland Pkwy, Las Vegas, NV 89154-4002, USA \\
$^{4}$ Kavli IPMU (WPI), The University of Tokyo, 5-1-5 Kashiwanoha, Kashiwa, Chiba, 277-8583, Japan \\
$^{5}$ SUPA, Institute for Astronomy, University of Edinburgh, Royal Observatory, Edinburgh, EH9 3HJ, UK
}
 
\begin{document}

\date{Accepted ?; Received ??; in original form ???}

\pagerange{\pageref{firstpage}--\pageref{lastpage}} \pubyear{2008}

\maketitle

\label{firstpage}

%
%
\begin{abstract}
Recent observations have successfully detected \oiii $88.3\,{\rm \mu m}$ and \cii $157.6\,{\rm \mu m}$  lines from galaxies in the early Universe with the Atacama Large Millimeter Array (ALMA).
Combining cosmological hydrodynamic simulations and radiative transfer calculations, 
we present relations between the metal line emission and galaxy evolution at $z=6-15$. 
We find that galaxies during their  starburst phases have high \oiii luminosity of $\sim 10^{42}~\rm erg~s^{-1}$. 
Once supernova feedback quenches star formation, \oiii luminosities rapidly decrease and continue to be zero  for $\sim 100\,{\rm Myr}$.
The slope of the relation between $\log{(\rm SFR/\Msunyr)}$ and $\log{(L_{\rm [O_{\III}]}/\Lsun)}$ at $z=6-9$ is 1.03, and 1.43 for $\log{(L_{\rm [C_{\II}]}/\Lsun)}$.
As gas metallicity increases from sub-solar to solar metallicity by  metal enrichment from star formation and feedback, 
the line luminosity ratio $L_{\rm [O_{\III}]} / L_{\rm [C_{\II}]}$ decreases from $\sim 10$ to $\sim 1$ because the  O/C abundance ratio decreases due to carbon-rich winds from AGB stars and the mass ratio of {\sc H\,ii} to {\sc H\,i} regions decreases due to rapid recombination.
Therefore, we suggest that the combination of \oiii and \cii lines 
is a good probe to investigate the relative distribution of ionized and neutral gas in high-$z$ galaxies. 
In addition, we show that deep \cii observations
with a sensitivity of $\sim 10^{-2}~{\rm mJy~arcsec^{-2}}$ can probe the extended neutral gas disks of high-$z$ galaxies. 
\end{abstract}

%
%


%
%

\section{Introduction}
\label{intro}
Understanding the physical properties of distant galaxies is one of the major goals of present astrophysical research. 
Observations of metal lines and dust continuum can be a useful tool to investigate the physical state of the ISM in distant galaxies. 
Recent ALMA observations have detected metal and dust in galaxies at $z\gtrsim 6$ via the \cii $158\,{\rm \mu m}$ line \citep{Willott15,Carniani17,Decarli17,Knudsen17,Smit18,Marrone18,Hashimoto19a}, the \oiii $88\,{\rm \mu m}$ line \citep{Inoue16,Carniani17,Laporte17,Hashimoto18a,Marrone18,Hashimoto19a,Tamura19}, and dust continuum \citep{Watson15,Laporte17,Bowler18,Marrone18,Hashimoto19a,Tamura19}. 
These observations can provide information about the physical state and star formation activities of galaxies.
For example, the \cii line is the main coolant for the warm ISM \citep{Wolfire03}, and leads to the formation of the cold neutral medium (CNM) via thermal instabilities even in a low-metallicity environment \citep{Arata18}. 
The confined CNM becomes molecular clouds, and forms stars. 
UV radiation from young stars subsequently produces {\sc O\,i\hspace{-.1em}i\hspace{-.1em}i} regions.
Therefore, combining \cii and \oiii observations will help to understand the physical processes of star formation within a multi-phase ISM \citep[e.g.][]{Cormier12}.

Recent observations showed interesting features in high-$z$ galaxies. 
\citet{Inoue16} reported \oiii detection for an Lyman-$\alpha$ emitter at $z=7.2$, but \cii was undetected. 
The upper limit of \cii luminosity was much lower than the expected from the relation of local galaxies \citep{DeLooze14}.
\citet{Laporte19} also reported the \cii absence for two \oiii emitters at $z \sim 9$ \citep[see also, ][]{Laporte17,Hashimoto18a}.
Meanwhile, \citet{Hashimoto19a} successfully detected both metal lines \cii and \oiii \citep[see also,][]{Marrone18,Smit18,Tamura19}. These galaxies show a negative correlation between the luminosity ratio $L_{\rm [O_{\III}]}/L_{\rm [C_{\II}]}$ and bolometric luminosity $L_{\rm bol}$.
The origin of these features has been a puzzle so far. 

Some theoretical works studied metal emission lines from high-$z$ galaxies. 
Combining cosmological simulations and analytical model for the multi-phase ISM, \citet{Nagamine06} predicted the ALMA observability of the \cii line from star-forming galaxies at $z=3-6$.
\citet{Pallottini17a} focused on properties of Lyman-Break Galaxies at $z\sim 6$.
They used zoom-in simulations, and found that H$_{2}$ galactic disk mainly contributed to the total \cii luminosity \citep[see also][]{Pallottini17b,Pallottini19}.
Recently, \citet{Moriwaki18} studied the observability of \oiii emitters at $z\sim 8$ using a $(50\,h^{-1}{\rm cMpc})^{3}$ simulation with a sub-grid model for the ionization structure.
Very recently \citet{Katz19} studied various metal emission lines at $z\gtrsim 9$ using zoom-in radiative hydrodynamics simulations and {\sc Cloudy}, and found kpc-scale offsets between \oiii and \cii emitting regions. 
However, their galaxy sample was limited due to the expensive calculation. 

As stated above, observed high-$z$ galaxies were likely to have a  wide variety of metal line properties. In this  work, we calculate the metal line properties of 10 galaxies with zoom-in initial conditions and study the relation between metal lines and galaxy evolution.  
As suggested in the simulations of \citet{Yajima17} (hereafter Y17), high-$z$ galaxies could repeat star-bursts, which results in galactic outflow due to SN feedback. This in turn quenches subsequent star formation, and results in intermittent star-formation histories \citep[see also,][]{Hopkins14,Kimm14,Davis14}.

Combining hydrodynamic simulations of Y17 and radiative transfer calculations, \citet[hereafter A19]{Arata19} studied how the intermittent star formation affected UV and infrared SEDs.
We found that the escape fraction of UV photons fluctuated due to changing dust distribution, which resulted in the fluctuations of IR luminosity.
In this paper, we focus on how FIR metal emission lines (\oiii 88\,${\rm \mu m}$ and \cii 158\,${\rm \mu m}$) are affected by the intermittent star formation.

Our paper is organized as follows.
We describe our models for cosmological hydrodynamic simulations and radiative transfer calculations in Section\,\ref{sec:method}.
In Section\,\ref{sec:result}, we present our results.
In Sections\,\ref{sec:map} and \ref{sec:phase}, we focus on which gas phases contribute to total metal line luminosities ($L_{\rm line}$).
In Section\,\ref{sec:redshift}, we focus on the fluctuations of $L_{\rm line}$ due to intermittent star formation in first galaxies.
We also show the relation between SFR and $L_{\rm line}$, and discuss how $L_{\rm [O_{\,\III}]}/L_{\rm [C_{\,\II}]}$ ratio decreases with metal enrichment.
In Section\,\ref{sec:line_ir}, we discuss the $L_{\rm line}/L_{\rm IR}$ ratio, which reflects the ratio of absorbed energy by gas or dust. 
In Section\,\ref{sec:dist}, we focus on the spatial distributions of metal emission lines.
In addition, we discuss the dependence of our results on sub-grid models for star formation and SN feedback in Sec.\,\ref{sec:model}.
Finally, we summarise our main conclusions in Section\,\ref{sec:summary}.


%
%
\section{Method}
\label{sec:method}

\subsection{Cosmological hydrodynamic simulations}
\label{sec:hydro}

We use the {\sc Gadget-3} code \citep[an updated version of {\sc Gadget-2} described in][]{Springel05a} with sub-grid models developed in the {\it Overwhelming Large Simulation} project \citep{Schaye10} and extended for the  {\it First Billion Year} (FiBY) project to include sub-grid models for galaxy formation in the early Universe such as e.g. POP-III star formation and molecular networks \citep[see e.g.][]{Johnson13,Paardekooper15}. 
The FiBY implementation reproduces general properties of the high-$z$ galaxy population well \citep[see e.g.][]{Cullen17,Arata19}.
We first focus on the two haloes presented in Y17, whose total masses are $M_{\rm h} \sim 10^{11}\,\Msun$ and $\sim 10^{12}\,\Msun$ at $z=6$ (Halo-11 and Halo-12), and call them the {\it fiducial} runs.
Halo-12 is identified as the most massive halo at $z\approx 3.0$ 
with $M_{\rm h} \sim 1.3\times 10^{13}\,h^{-1}\Msun$ in a simulation box of comoving $(100\,{\rm Mpc})^{3}$. 
To increase our simulated galaxy samples, we perform additional cosmological hydrodynamic simulations.
Here we briefly summarize our models.

\begin{figure}
\begin{center}
\includegraphics[width=\columnwidth]{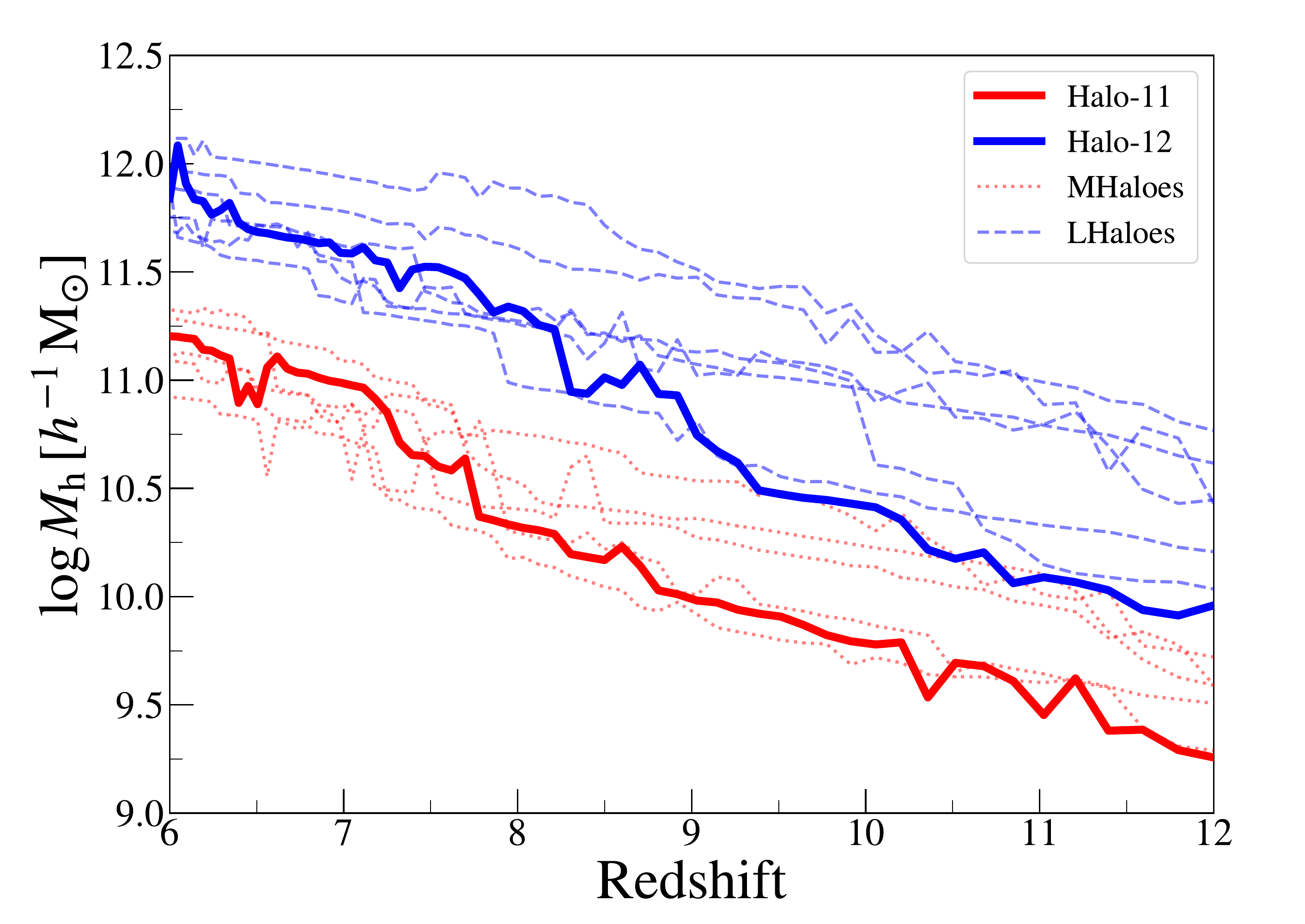}
\caption{Mass growth of our haloes in the zoom-in simulations. Red lines represent haloes with total halo masses of $M_{\rm h} \sim 10^{11}\,\Msun$ at $z=6$, identified in $(20\,{\rm cMpc})^{3}$ box. Blue lines are for haloes with $M_{\rm h} \sim 10^{12}\,\Msun$ at $z=6$, identified in $(100\,{\rm cMpc})^{3}$ box. Thick lines represent our fiducial runs (Halo-11 and Halo-12). Table\,\ref{table:setup} shows detailed properties of these haloes. 
}
\label{fig:halos}
\end{center}
\end{figure}

\begin{table*}
  \centering
  \begin{tabular}{cccccccc}
    \hline
    Halo ID  & $\un{M}{h}~[h^{-1}~\Msun]$  &  $\un{m}{DM}~[h^{-1}~\Msun]$  &  $\un{m}{gas}~[h^{-1}~\Msun]$  &  SNe feedback  &  $A$ & $\delta$ & $M_{\rm 1500}\,{\rm [mag]}$\\
    \hline \hline
    Halo-11  & $1.6\times 10^{11}$  & $6.6\times 10^{4}$ & $1.2\times 10^{4}$ & ON & $2.5\times 10^{-3}$ & 1.52 & $-20.3$\\
    Halo-12 & $6.9\times 10^{11}$  & $1.1\times 10^{6}$ & $1.8\times 10^{5}$ & ON & $2.5\times 10^{-3}$ & 2.02 & $-22.1$\\
	\hline
    MHalo-0  & $2.1\times 10^{11}$  & $6.6\times 10^{4}$ & $1.2\times 10^{4}$ & ON & $2.5\times 10^{-3}$ & 1.54 & $-20.4$\\    
	MHalo-1  & $1.9\times 10^{11}$  & $6.6\times 10^{4}$ & $1.2\times 10^{4}$ & ON & $2.5\times 10^{-3}$ & 1.48 & $-20.4$\\
    MHalo-2  & $1.3\times 10^{11}$  & $6.6\times 10^{4}$ & $1.2\times 10^{4}$ & ON & $2.5\times 10^{-3}$ & 1.66 & $-20.2$\\
    MHalo-3  & $1.2\times 10^{11}$  & $6.6\times 10^{4}$ & $1.2\times 10^{4}$ & ON & $2.5\times 10^{-3}$ & 1.58 & $-20.3$\\    
    \hline 
    LHalo-0  & $1.3\times 10^{12}$  & $1.0\times 10^{6}$ & $1.8\times 10^{5}$ & ON & $2.5\times 10^{-3}$ & 3.12 & $-23.2$\\    
    LHalo-1  & $9.3\times 10^{11}$  & $1.0\times 10^{6}$ & $1.8\times 10^{5}$ & ON & $2.5\times 10^{-3}$ & 2.50 & $-22.5$\\
    LHalo-2  & $7.9\times 10^{11}$  & $1.0\times 10^{6}$ & $1.8\times 10^{5}$ & ON & $2.5\times 10^{-3}$ & 2.29 & $-22.0$\\
    LHalo-3  & $5.7\times 10^{11}$  & $1.0\times 10^{6}$ & $1.8\times 10^{5}$ & ON & $2.5\times 10^{-3}$ & 2.20 & $-21.0$\\    
    LHalo-4  & $8.0\times 10^{11}$  & $1.0\times 10^{6}$ & $1.8\times 10^{5}$ & ON & $2.5\times 10^{-3}$ & 2.19 & $-20.3$\\
    LHalo-5  & $4.9\times 10^{11}$  & $1.0\times 10^{6}$ & $1.8\times 10^{5}$ & ON & $2.5\times 10^{-3}$ & 2.09 & $-21.2$\\
    \hline
    Halo-11-lowSF  & $1.5\times 10^{11}$  & $6.6\times 10^{4}$ & $1.2\times 10^{4}$ & ON & $2.5\times 10^{-4}$ & 1.52 & $-19.5$\\ 
    Halo-11-noSN  & $1.6\times 10^{11}$  & $6.6\times 10^{4}$ & $1.2\times 10^{4}$ & OFF & $2.5\times 10^{-3}$ & 1.52 & $-21.0$\\ 
	Halo-12-lowSF  & $8.0\times 10^{11}$  & $1.1\times 10^{6}$ & $1.8\times 10^{5}$ & ON & $2.5\times 10^{-4}$ & 2.02 & $-21.7$\\
	Halo-12-noSN  & $6.5\times 10^{11}$  & $1.1\times 10^{6}$ & $1.8\times 10^{5}$ & OFF & $2.5\times 10^{-3}$ & 2.02 & $-22.4$\\
    \hline
  \end{tabular}
\caption{Parameters of our zoom-in cosmological hydrodynamic simulations: (1) $\un{M}{h}$ is the halo mass identified by the friends-of-friends method for dark matter particles at $z = 6.0$. (2) $\un{m}{DM}$ is the mass of a dark matter particle. (3) $\un{m}{gas}$ is the initial mass of a gas particle. 
(4) $A$ is the amplitude factor in the star formation model based on the Kennicutt--Schmidt law \citep{Schaye08}.
The Halo-11-lowSF and Halo-12-lowSF runs have a lower star formation amplitude factor. The Halo-11-noSN and Halo-12-noSN runs have no SN feedback. 
(5) $\delta \equiv \rho/\rho_{\rm crit}$ is the mass overdensity at $z=10$ within a sphere of radius $1\,{\rm cMpc}$ centered on the halo.
(6) $M_{1500}$ is the absolute magnitude at $z=6.0$ in rest-frame $1500\,$\AA,~obtained from radiative transfer. 
}
\label{table:setup}
\end{table*}

We first conduct $N$-body simulations with box-sizes of $(20\,h^{-1}{\rm Mpc})^{3}$ and $(100\,h^{-1}{\rm Mpc})^{3}$, and identify the four most massive haloes at $z=6$.
We call them MHalo-0, 1, 2, 3 and LHalo-0, 1, 2, 3, respectively.
We also perform an additional N-body simulation with $(100\,h^{-1}{\rm Mpc})^{3}$ box size until $z=3$ and identify the two most massive haloes with masses of $1.8\times 10^{13}\,h^{-1}\Msun$ and $1.3\times 10^{13}\,h^{-1}\Msun$.  We then recompute hydrodynamics in the zoom-in region until $z=6$ (LHalo-4, 5).
We set the gravitational softening length to $\epsilon_{\rm min} = 200\,{\rm pc}$ in comoving units for every zoom runs.

Figure\,\ref{fig:halos} shows the redshift evolution of halo masses in the zoom runs.
We observe differences in the growth rates between Halo-12 and the LHaloes, which we ascribe to the environments that they live in. 
This can be quantified by the mass overdensity $\delta$ at $z= 10$ within a sphere of radius $1\,{\rm cMpc}$ centered on each halo: $\delta=2.02$ (for Halo-12), and $\delta = 3.12,~2.50,~2.29,~2.20,~2.19$ and $2.09$ for LHalo-0, 1, 2, 3, 4 and 5 respectively.  
Here, we measure the overdensity as $\delta \equiv \frac{\left< \rho \right> - \overline{\rho}}{\overline{\rho}}$ with a spherical top-hat window of comoving radius 1\,Mpc, where $\left< \rho \right>$ and $\overline{\rho}$ are the mean matter density in the zoom-in regions and the cosmic mean density, respectively.
The overdensity $\delta$ for LHalo-0, 1, 2 and 3 corresponds to $1.57 \sigma_{1\,{\rm Mpc}}$.
From these values, it is evident that LHaloes live in more dense environments than Halo-12, which results in earlier growth of halo masses.
At $z\sim 6$, Halo-12 catches up with LHaloes and achieves an intermediate halo mass among them.
Also, Halo-11 shows somewhat slower growth than MHalo-0, 1, 2 with higher $\sigma$ values. In practice, the halo growth at $z \sim 10$ could be affected by  smaller-scale density fluctuation, and may not be tightly correlated with  the overdensity on 1\,Mpc scale.  Nevertheless, the different overdensity within 1\,Mpc do reflect  the different growth rates of haloes roughly.

Next we make zoom-in initial conditions with effective resolution of $2048^{3}$ particles for $(20\,h^{-1}{\rm Mpc})^{3}$ box and $4096^{3}$ particles for $(100\,h^{-1}{\rm Mpc})^{3}$ box using the {\sc MUSIC} code \citep{Hahn11}, and carry out hydrodynamics simulations up to $z=6$.
The detailed information of the models is shown in Table~\ref{table:setup}.

The local star formation rate is calculated as \citep{Schaye08}
\begin{equation}
\dot{m}_{*} =  A (1\,\Msun~{\rm pc^{-2}})^{-n} m_{\rm g} 
\left(
\frac{\gamma}{G} f_{\rm g} P_{\rm tot}
\right)^{(n-1)/2},
\end{equation}
where $\gamma$ is a specific heat index, and $m_{\rm g}$ is the mass of gas particle.
The $f_{\rm g}$ is gas mass fraction in the self-gravitating galactic disk, and $P_{\rm tot}$ is total ISM pressure.
This equation is based on the Kennicutt--Schmidt relation 
$\dot{\Sigma}_{*} = A (\Sigma_{\rm g}/1\,\Msun~{\rm pc^{-2}})^{n}$ \citep{Kennicutt98}.
We assume $f_{\rm g} = 1$ and $n=1.4$.
We also use $A=2.5 \times 10^{-3}\,\Msun~{\rm yr^{-1}~{\rm kpc^{-2}}}$ as the fiducial value, which is 10 times higher than that of local star-forming galaxies, but recent observations have shown that higher values are preferred for merging or high-$z$ galaxies \citep[e.g.][]{Genzel10,Tacconi13}.
To check the impact of $A$ on our main results,
we also study a run with $A=2.5 \times 10^{-4}\,\Msun~{\rm yr^{-1}~{\rm kpc^{-2}}}$ (indicated as `low-SF' in Table\,\ref{table:setup}).
Our star formation threshold density is $n_{\rm H} = 10\,\cc$  \citep{Johnson13}.

Our simulations track the abundances of 9 elements (H, He, C, N, O, Ne, Mg, Si and Fe) for each particle separately \citep{Wiersma09b}.
The sources of chemical enrichment are Type-Ia/II  SNe, and AGB stars \citep{Portinari98,Marigo01}.
For $z\gtrsim 6$ galaxies, it is dominated by Type-II SNe, because it takes $\sim 10^{9}\,{\rm yrs}$ to become AGB stars for low- and intermediate-mass stars. 
However, as we will show later, the AGB star contribution is important for some of the abundance ratio, such as O/C.
In this paper, we consider all of the above chemical sources in our  calculations of metal line luminosities  (Sec.\,\ref{sec:oiii}, \ref{sec:cii}).

SN feedback injects thermal energy into neighbouring gas particles stochastically \citep{Dalla12}. 
The injection energy is $10^{51}\,{\rm erg}$ per single SN event, and the randomly selected gas particles are heated up to $10^{7.5}\,{\rm K}$.
The thermal energy is  efficiently converted into kinetic energy against radiative cooling, if the gas density is lower than the critical value of $n_{\rm H} \sim 100\,\cc\,(T/10^{7.5}\,{\rm K})^{3/2} (m_{\rm g}/10^{4}\,\Msun)^{-1/2}$.
The numerical resolution of our simulations are adequate to successfully launch galactic winds and suppress star formation (see Y17).
We further perform simulations without SN feedback to investigate the impact on metal line properties (indicated as `no-SN' in Table\,\ref{table:setup}).
In this case, surrounding gas is not heated by the SN feedback, but only the metal enrichment occurs into neighbouring gas particles.

%
%
\subsection{Radiative transfer}
\label{sec:rt}

In next sub-sections, we model metal emissions based on the ionization structure of hydrogen gas.
To obtain it, we use the radiative transfer code, All-wavelength Radiative Transfer with Adaptive Refinement Tree ({\sc Art$^2$}) code \citep{Li08, Yajima12a}.
The details of this code were described in A19.

The {\sc Art$^2$} is based on the Monte Carlo technique. 
It tracks propagation of photon packets emitted from stellar particles, and computes the emergent SED.
We use a total of $2 \times 10^{5}$ photon packets for ionizing ($>13.6$ eV) and non-ionizing ($<13.6$ eV) radiation with 500 frequency bins.

The intrinsic SED for each stellar particle is taken from {\sc Starburst99} \citep{Leitherer99} assuming the Chabrier initial mass function with a  mass range of $0.1-100\,\Msun$ \citep{Chabrier03}.
{\sc Art$^2$} calculates the transfer of ionizing photons, UV continuum and dust absorption/re-emission using adaptive mesh refinement (AMR) cells.
The spatial resolution of the minimum size cells is set to physical $2.7\,h^{-1}{\rm pc}$ for Halo-11 at $z=6$ which is similar to the minimum smoothing length of SPH particles.
The construction of the AMR cells is based on the gas density structure. The minimum smoothing length of gas particles is 20 comoving pc, and the physical scale becomes close to the minimum cell size $2.7\,h^{-1}$\,pc at $z>6$. Thus our AMR grid follows the detailed ISM structures which are described by the SPH simulations. The stellar distribution is restricted by the softening length of 200 comoving pc, which brings uncertainties in the small scale structures of {\sc Hii} regions ($\lesssim 30$ physical pc). However, our main interests are in the total metal-line luminosities and large scale morphologies ($\sim 10$ physical kpc). Therefore, our results should not change significantly due to the uncertainties of the super-sampling by the AMR grid.

We assume that the dust-to-gas mass ratio $\mathcal{D}$ is proportional to gas metallicity as seen in local galaxies, $\mathcal{D} = 8\times 10^{-3}\,(Z/\Zsun)$ \citep{Draine07}.
We adopt the dust size distribution derived in \citet{Todini01}.
They modeled the formation and evolution processes of dust grains in the expanding ejecta of SN, and investigated the size distribution.
More massive SN progenitors create heavy elements more efficiently, however, large amounts of metals in the internal layers fall back to the center if the kinetic energy is not large enough for escape. Thus they argue that intermediate-mass progenitors are the main sources of interstellar dust.
We use their model for SN from stars of with  $M=22\,\Msun$ and solar metallicity.
For example, this model could explain the extinction curves of $z\sim 6$ quasars   \citep{Maiolino04}. Also, observability of high-$z$ dust emission by the ALMA was reproduced with large simulation samples \citep{Arata19}.
This model uses the cross-sections for absorption and scattering \citep{Weingartner01} to obtain the dust opacity.
In addition, before the radiative transfer calculation, the local $\mathcal{D}$ is multiplied by a scaling factor which is a function of gas temperature of the cell, $C = 1-(1-\mathcal{D}_{\rm min}) \exp{(-(3000/T)^2)}$, where $\mathcal{D}_{\rm min}=0.01$ is the minimum dust-to-gas mass ratio, considering that the dust abundance is much lower in the ionized regions \citep{Burke74,Reynolds97}. Note that we focus on high-redshift galaxies whose ISM is not heavily enriched by metals. Therefore, the dust absorption in ionized regions is unlikely to be significant.

In addition, we do not consider the UV background in our RT calculations, which is important for the ionization state of the IGM \citep{Haardt12}. However, we are mainly interested in gas at densities higher than that for self-shielding \citep[$n\sim 10^{-2}\,\cc$;][]{Nagamine10b,Altay11,Yajima12b,Bird13,Rahmati13}. 
Thus our conclusions do not change by neglecting the effect of the UVB.

%
%
\subsection{\oiii emission model}
\label{sec:oiii}

UV photons from massive stars create \oiii regions.
Therefore, the \oiii luminosity is likely to be related closely with star formation activity. 
Recent observations successfully detected \oiii lines and confirmed the redshifts of distant galaxies even at $z \gtrsim 7$ \citep[e.g.,][]{Hashimoto18a}. 
Since the ionization potential of O$^{+}$ $\to$ O$^{2+}$ is $35.121\,{\rm eV}$,  O$^{2+}$ ions exist only in {\sc H\,ii} regions.
Based on the photo-ionization radiative transfer calculations (Sec. \ref{sec:rt}), we classify all cells into `{\sc H\,i} cell' and `{\sc H\,ii} cell' according to their hydrogen neutral fractions, separated at the value of 0.5. 
Next we calculate doubly ionized fraction of oxygen in each {\sc H\,ii} cell assuming ionization equilibrium between O$^{+}$ and O$^{2+}$ \footnote{Here we neglect the {\sc O\,i} abundance, because the ionization potential is similar to that of {\sc H\,i} ($\Phi_{0} = 13.6181\,{\rm eV}$), and most {\sc O\,i} atoms are ionized in {\sc H\,ii} region.}:
\begin{equation}
\label{eq:ioneq}
 \int_{\nu_{\rm min}}^{\infty} n_{\rm O^{+}}
\frac{\sigma_{\rm \nu} F_{\rm \nu}}{h\nu} d\nu
=
 \alpha(T) n_{\rm e} n_{\rm O^{2+}}.
\end{equation}
The left hand side represents the photo-ionization rate per volume, where $\nu_{\rm min}=8.492\times 10^{15}\,{\rm Hz}$ is the minimum frequency for ionization.
For the photo-ionization cross section $\sigma_{\rm \nu}$  we use the fitting function derived in \citet{Verner96}  \citep[same as in {\sc Cloudy},][]{Ferland98,Richings2014}.
We evaluate the flux $F_{\nu}$ using the optically-thin approximation and sum up contributions from nearby stellar particles:
\begin{equation}
F_{\rm \nu} = \sum_{R_{\rm j}<R_{\rm S}} F_{\rm \nu,j} = \sum_{R_{\rm j}<R_{\rm S}} \frac{L_{\rm \nu,j}}{4\pi R_{\rm j}^{2}},
\label{eq:incomingflux}
\end{equation}
where $L_{\rm \nu,j}$ is the specific luminosity of the $j$-th stellar particle, and $R_{\rm j}$ is the distance from the cell.
In the above estimation, we calculate the radius of the Str\"{o}mgren sphere as $R_{\rm S} = \left( 3Q_{0} / 4 \pi n_{\rm H}^{2} \alpha_{\rm B} \right)^{1/3}$, where $Q_{0}$ [$s^{-1}$] is the ionizing photon emissivity of a nearby stellar particle.
If the radius reaches a target cell, the flux from stars in the nearby cell is considered using the above equation. 
We need this condition to exclude the contributions from stellar particles in the other {\sc H\,ii} regions, because their UV radiation should be interrupted by intervening neutral gas.
In this estimate, we do not consider the expansion of the Str\"{o}mgren sphere due to the overlap of ionizing flux from nearby stars. Also, the dust absorption is ignored here because of the modeled low dust content in {\sc H\,ii} regions.
The right-hand-side of Eq.\,(\ref{eq:ioneq}) represents recombination rate per volume, where $\alpha(T)$ is the coefficient as a function of temperature \citep{Nahar97,Nahar99}.

The  O$^{2+}$ ground state has three fine-structures: ${\rm ^{3}P_{0}}$, ${\rm ^{3}P_{1}}$, and ${\rm ^{3}P_{2}}$.
The $88\,{\rm \mu m}$ FIR line is emitted via the ${\rm ^{3}P_{1}}$ $\to$ ${\rm ^{3}P_{0}}$ transition. 
We calculate the rate equations among three levels \citep{Nussbaumer81}, and obtain the level populations.
Local \oiii luminosity from a {\sc H\,ii} cell is estimated by
\begin{equation}
\label{eq:lumi}
L = (C_{\ell u}n_{\ell} - C_{u\ell}n_{\rm u}) \beta h\nu_{u\ell}V_{\rm cell},
\end{equation}
where $C_{\ell u}$ ($C_{u\ell}$) is the Einstein coefficient of collisional excitation (de-excitation) which depends on electron density. 
The $V_{\rm cell}$ is the cell volume, and 
$\beta$ is the escape probability, for which we assume optically thin case of $\beta=1$.
In addition, if gas temperature is higher than $1.2\times 10^{5}\,{\rm K}$, we set the luminosity to zero because the ionization degree of oxygen is triple or higher \citep{Nahar99}.
We show the comparison between our model and {\sc Cloudy} in Appendix \ref{sec:cloudy}. Most of the gas shows \oiii emission similar to {\sc Cloudy}, while a part of the gas deviates from it by a factor of a few. However, we find that the total \oiii luminosity is almost the same as the {\sc Cloudy} result.

%
%

\subsection{\cii emission model}
\label{sec:cii}

The \cii $158\,{\rm \mu m}$ FIR line can be radiated from gas in various phases, i.e., warm neutral and ionized medium, CNM and molecular clouds \citep[e.g.][]{Wolfire03}, thus the modeling is relatively difficult.
Previous theoretical work studied the contributions from each component with cosmological hydrodynamics simulations \citep{Olsen17,Pallottini17b,Katz19}. 
\citet{Olsen17} showed that diffuse ionized gas and molecular clouds were the main contributors to the \cii luminosity.
Whereas \citet{Pallottini17b} and \citet{Katz19} suggested that the \cii luminosity was connected to gas in dense environment.
Observational constrains on the contributions suffer from the difficulty of determining the optical depth to the line \citep[e.g.][]{Neri14}. 

The \cii emission model is similar to that of \oiii emission described in previous sub-section. However, in the case of \cii, we consider both  {\sc H\,i} and {\sc H\,ii} cells as the emission sites. 
For the {\sc H\,ii} cells, we calculate the ionization equilibrium between C$^{+}$ and C$^{2+}$ under the stellar radiation field (Eq.\,\ref{eq:ioneq}).
Meanwhile, we assume that all carbons are in C$^{+}$ ions for {\sc H\,i} cells. 
In our Galaxy, carbons are almost completely in C$^{+}$ ions under the FUV  ($G_{0} = 1.0\,{\rm Habing}$) and cosmic ray background which are radiated by nearby star-forming regions and external galaxies \citep[e.g.][]{Webber98,Seon11}.
Thus our assumption is valid 
if the galaxy is exposed to such a strong radiation field in the Habing band,
or at least our model provides an upper limit of the \cii luminosity.
If we consider the attenuation of the Habing band by hydrogen molecules, the C$^{+}$ abundance could decrease because some fraction of carbons would be {\sc C\,i}.

The \cii 158\,${\rm \mu m}$ line is radiated via the transition  ${\rm ^{2}P_{3/2}}$ $\to$ ${\rm ^{2}P_{1/2}}$.
The luminosity is calculated as in Eq.\,\ref{eq:lumi}. 
If the temperature of a cell is higher than $4.0\times 10^{4}\,{\rm K}$, the luminosity  is set to zero because carbons are collisionally ionized to double or higher \citep{Nahar97}.
Here we adopt the optically-thin approximation for the \cii 158\,${\rm \mu m}$ line, i.e., $\beta=1$. 
Note that, however, \citet{Neri14} suggested that the optical depth for an observed $z\sim 5.2$ galaxy could be $\tau_{\rm [C\,II]} \gtrsim 1$.
If this is common in high-$z$ galaxies, our model may somewhat overestimate the \cii luminosity. 
In addition, we also note that the cosmic microwave background (CMB) can affect \cii emission \citep[e.g.][]{Goldsmith12,daChunha13}.
As the redshift increases, the CMB temperature ($T_{\rm CMB}=2.73(1+z)\,{\rm K}$) approaches  the equivalent temperature of \cii emission ($T_{\rm eq}=91.2\,{\rm K}$), although it is still smaller by a factor of a few. Thus stimulated emission/absorption may be significant for $z \gtrsim 6$ galaxies \citep{Vallini15,Lagache18}.
In Appendix \ref{sec:cmb}, we estimate the CMB effect and 
find that the stimulated emission/absorption due to CMB (spin temperature coupling) is not significant. Therefore, the CMB reduces the luminosity only by a few percent.


%
%

\section{Result} 
\label{sec:result}


\subsection{Projected Images}
\label{sec:map}

\begin{figure*}
\begin{center}
\includegraphics[width=2\columnwidth]{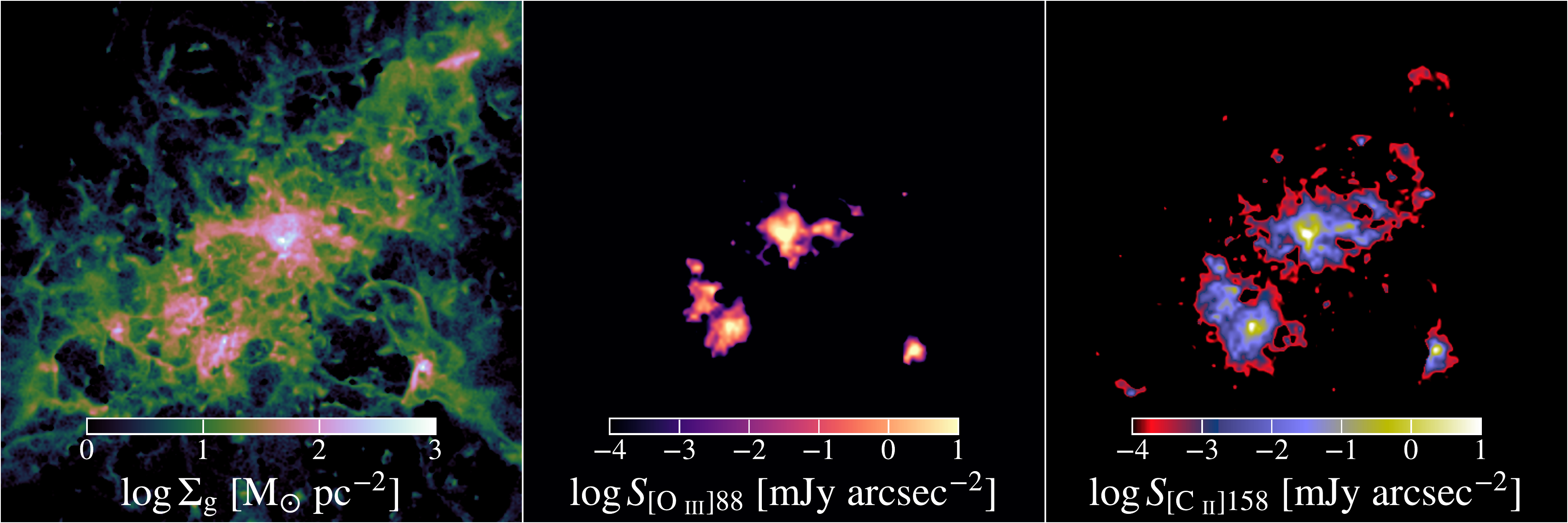}
\caption{Maps of the main halo in Halo-11 run at $z=6.0$. {\it Left panel}: Gas surface density. The spacial scale is $51$ physical kpc. {\it Middle \& right panels}: Intensity maps of \oiii 88\,${\rm \mu m}$ and \cii 158\,${\rm \mu m}$ lines, respectively. The pixel size is $\sim 0.07\,{\rm arcsec}$, which corresponds to $\sim 0.4$ physical kpc. 
}
\label{fig:maps}
\end{center}
\end{figure*}

\begin{figure*}
\begin{center}
\includegraphics[width=2\columnwidth]{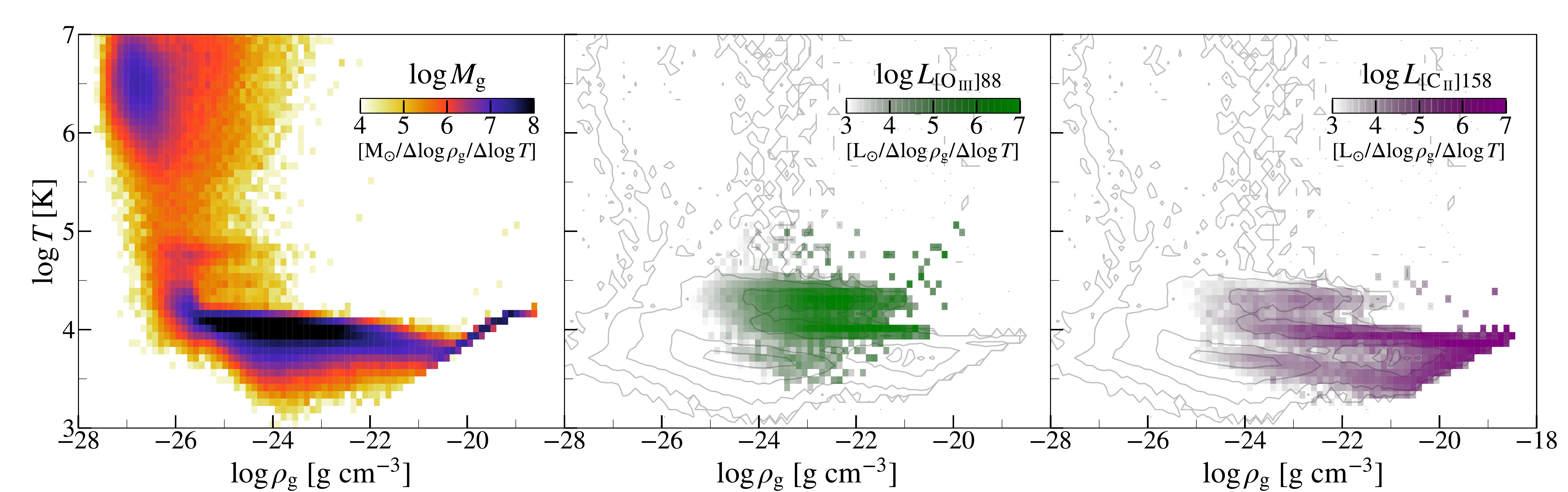}
\caption{{\it Left panel}: Gas phase diagram for Halo-11 at $z = 6.0$. {\it Middle \& right panels}: Gas phases emitting \oiii 88\,${\rm \mu m}$ and \cii 158\,${\rm \mu m}$ lines, respectively. The gray contour shows gas phases after the ionization radiative transfer calculation, where five levels of contour correspond to   the following gas mass in each pixel: $\log{M_{\rm g}\,[\Msun / \Delta \log{\rho_{\rm g}} / \Delta \log{T}]} = 4,~5,~6,~7$ \& 8.
}
\label{fig:phase}
\end{center}
\end{figure*}

The left panel of Figure\,\ref{fig:maps} presents a 2-D map of gas column density in Halo-11 at $z=6.0$.
The total stellar mass, gas mass, and dust mass are $M_{\rm \star} = 2.2\times 10^{9}\,h^{-1}\Msun$, $M_{\rm gas} = 1.1\times 10^{10}\,h^{-1}\Msun$ and $M_{\rm dust} = 9.6\times 10^{6}\,h^{-1}\Msun$.
The gas structure shows filaments and clumps, 
and extends over $\sim 20\,{\rm kpc}$.
The middle and right panels show the surface brightness of \oiii 88\,${\rm \mu m}$ and \cii 158\,${\rm \mu m}$, which
represent ionized and neutral regions (see Fig.\,\ref{sec:phase}), and the brightest pixels have 157.6 and 485.7\,${\rm mJy~arcsec^{-2}}$, respectively.
We find that the \oiii map is sharply cut off at high-density regions of $\Sigma_{\rm gas} \gtrsim 10^{2}\,{\rm \Msun~pc^{-2}}$ (white in the left panel). 
In addition,  future deep \cii observations with a sensitivity of $\lesssim 10^{-2}\,{\rm mJy~arcsec^{-2}}$ will be able to probe the structure of extended cold gas disks over $\sim 10\,{\rm kpc}$ (pink in the left panel).
The integration time of $\sim 78$ hours with the fully operated ALMA achieves $5\sigma$ detection. We here use the ALMA sensitivity calculator\footnote{ https://almascience.eso.org/proposing/sensitivity-calculator}.


\subsection{Physical properties of gas with metal-line emission}
\label{sec:phase}

The left panel of Figure~\ref{fig:phase} shows the phase diagram of gas in Halo-11 at $z=6.0$.
The hot ionized medium ($T\sim 10^{6}-10^{7}\,{\rm K}$) cools down to $\sim 10^{4}\,{\rm K}$ via metal cooling \citep{Wiersma09a} and hydrogen $\lya$ cooling. Star formation occurs in dense gas clouds ($n_{\rm H}\gtrsim 10\,\cc$). The star-forming clouds contracts based on the effective equation of state \citep{Schaye08}.
Due to SN feedback, the gas near young stars is returned to the hot phase.
In this work, we consider the photo-ionization of gas due to young stars by post-processing, and heating up the gas to $\sim 2 \times 10^{4}\,{\rm K}$ if the ionization degree of hydrogen exceeds 0.99.
In the post-processing ionization transfer, the minimum temperature of gas is set to the results of the hydrodynamic simulations.
The corrected gas temperature is shown by the gray contours in the middle and right panels.

The middle panel of Fig.\,\ref{fig:phase} shows the physical state of gas emitting the \oiii line. 
The total luminosity is $2.26 \times 10^{42}\,{\rm erg~s^{-1}}$, and the half of it is contributed by completely ionized gas ($x_{\rm e} > 0.99$).
The right panel shows \cii emitting phases. 
The total luminosity is $2.0 \times 10^{42}\,{\rm erg~s^{-1}}$, and 
its 99\% is contributed by the neutral gas ($x_{\rm e} < 0.5$).  
In particular, the gas clouds with $n_{\rm H} \gtrsim 10^{3}~\rm cm^{-3}$ are the main contributor (88\%) to the total luminosity.
The large contribution from high-density regions is similar to the results of previous works \citep{Pallottini17a, Katz19}.
Since the density is higher than the critical density for thermalization of 158\,${\rm \mu m}$ fine-structure transition ($\sim 10^{3}\,\cc$, see Fig.\,\ref{fig:cmb}), the cooling rate per hydrogen atom is saturated.
Observationally, \citet{Croxall17} studied the contribution from neutral gas using local star-forming galaxies, and showed that it decreased from $\sim 0.9$ to $\sim 0.6$ as gas metallicity increased from $12+\log{\rm (O/H)}\sim 8.0$ to $8.6$.
The metallicity of Halo-11 at $z=6.0$ is $12+\log{\rm (O/H)}\approx 8.1$.
Thus the observational result supports our very high contribution from the neutral gas.
We show that the metal-line luminosities rapidly change due to SN feedback as shown in the next sub-section.


\subsection{Redshift Evolution of Metal Emission Lines}
\label{sec:redshift}

\begin{figure}
\begin{center}
\includegraphics[width=\columnwidth]{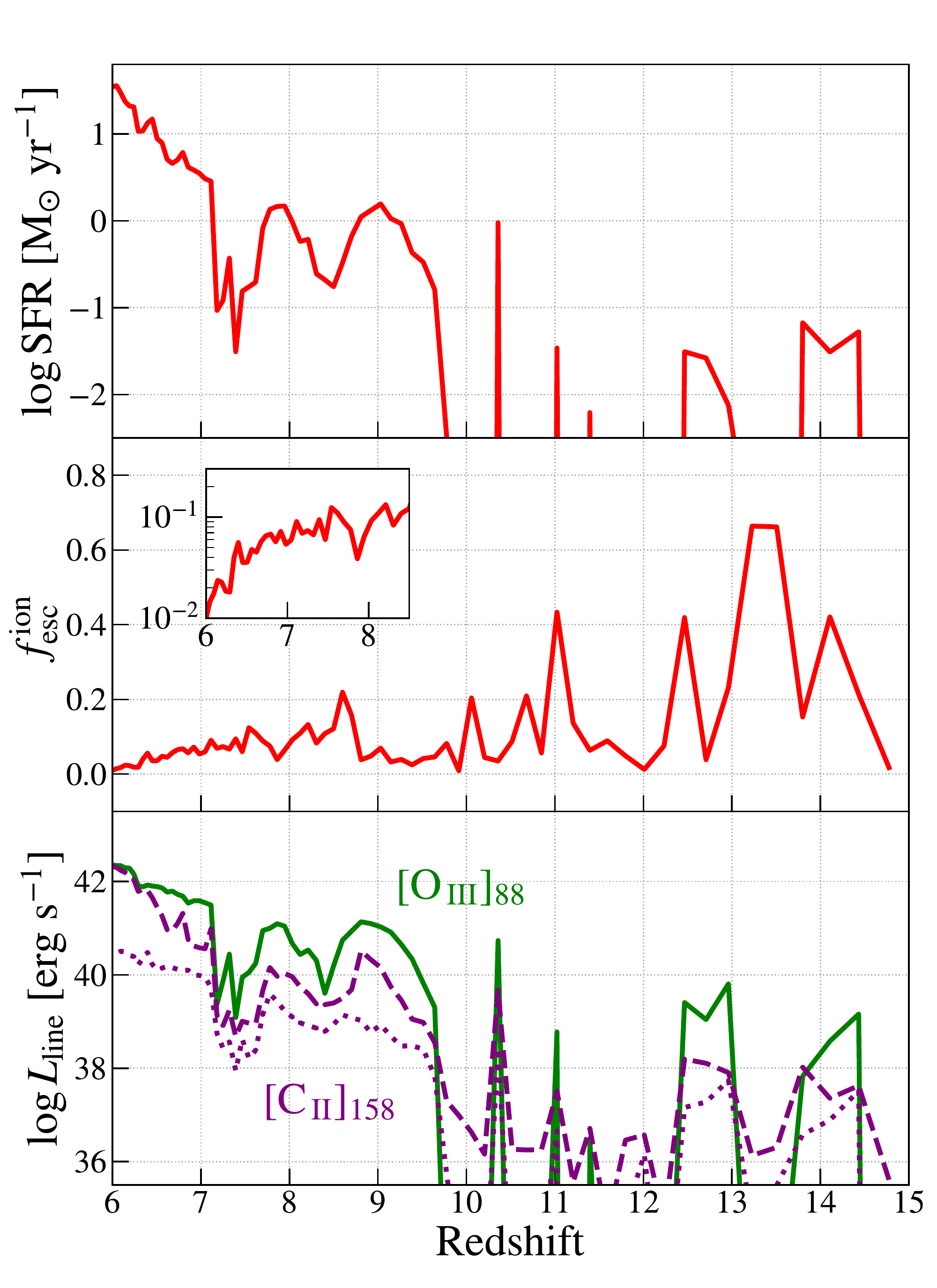}
\caption{Redshift evolution of SFR ({\it top}), escape fraction of ionizing photons ({\it middle}), metal-line luminosities ({\it bottom}) for Halo-11. In the middle panel, the inset shows the escape fraction at $z=6-8$ in logarithmic-scale. In the bottom panel, green solid line represents \oiii 88\,${\rm \mu m}$ luminosity. Purple dashed line is for \cii 158\,${\rm \mu m}$ luminosity.
Dotted line shows the contribution to \cii luminosity from {\sc H\,ii} regions.
}
\label{fig:halo11}
\end{center}
\end{figure}

As shown in A19, UV and IR continuum flux from high-$z$ galaxies rapidly change due to the intermittent star formation history and SN feedback. Here we study the time evolution of metal lines. 
The top panel of Figure\,\ref{fig:halo11} shows the redshift evolution of the SFR of Halo-11 at $z=6-15$.
SN feedback evacuates most of the gas in galaxies and quenches  star formation. The galaxy repeats the cycle of  starburst and quenching at $z \gtrsim 10$. The time-scale of fluctuation is $\sim 100 \left( \frac{1+z}{10} \right)\,{\rm Myr}$ which corresponds to the free-fall time of the halo.
As the halo mass increases, the deep gravitational potential holds the gas against SN feedback, resulting in more continuous star formation \citep[see][for details]{Arata19}.

The middle panel of Fig.\,\ref{fig:halo11} shows the escape fraction of ionizing (Lyman-continuum) photons.
During the star-burst phases, ionizing photons are efficiently absorbed by dusty clouds surrounding star-forming regions, while $f_{\rm esc}$ increases during outflow phases because SN feedback makes holes allowing photons to escape. 
This results in the fluctuation of $f_{\rm esc}$ between $\sim 0.01 - 0.6$ at $z>10$.
Recent simulations also showed a similar trend \citep{Paardekooper13,Kimm14, Kimm17,Katz18a}. 
\citet{Kimm14} showed that $f_{\rm esc}$ fluctuated in $0.01 - 0.9$ and the time-averaged $f_{\rm esc}$ decreased with increasing halo mass by using cosmological radiative hydrodynamic simulations.

\begin{figure}
\begin{center}
\includegraphics[width=\columnwidth]{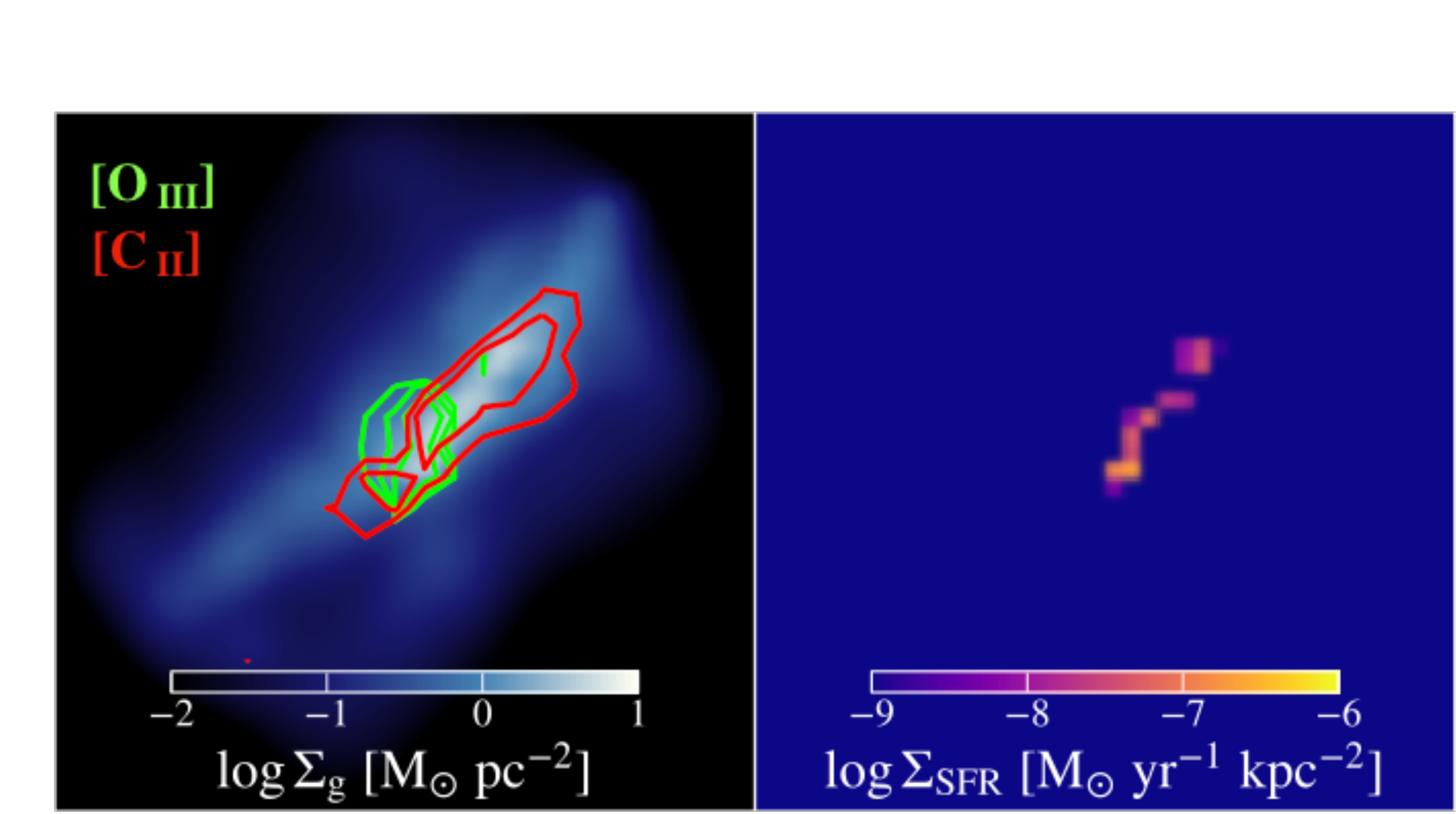}
\caption{
Maps of the main halo in Halo-11 run at $z=11$ with the spatial scale of $\sim 8$ physical kpc. {\it Left panel:} Gas surface density and contours of \oiii and \cii intensity maps. The contour lines represent $\log{S}~{\rm [\mu Jy~arcsec^{-2}]}$ of 1,~2~\&~3 for \oiii and  0~\&~1 for \cii. {\it Right panel:} Projected star formation rate density.
}
\label{fig:offset}
\end{center}
\end{figure}

At $z=11$, the gas structure becomes filamentary, and star formation occurs in the knots as shown in Fig.\,\ref{fig:offset}.
The ionizing photons efficiently escape into the perpendicular direction to the filament ($f_{\rm esc}\sim 40\,\%$).
As described in the previous sections, the \oiii and \cii lines are emitted from ionized regions (knots) and neutral gas (filament), respectively.
Thus we suggest that the combination of \oiii and \cii observations will reveal the neutral and ionized gas distribution, and indicate the escaping direction of ionizing photons. 
In addition, other simulations, which include radiative feedback from massive stars, have suggested rapid expansion of {\sc H\,ii} bubbles \citep[e.g.][]{Hopkins14}.  This implies extended low-density \oiii regions.
Meanwhile, our simulations predict very compact and high-density regions with $n\gtrsim 10\,\cc$, extending over $\sim 5\,{\rm physical~kpc}$ with a surface brightness of $\gtrsim 1\,{\rm mJy~arcsec^{-2}}$ at $z=6$ (see Fig.\,\ref{fig:maps}, the half light radius of the central clump is $\lesssim 1\,{\rm kpc}$ as discussed in Sec.\,\ref{sec:dist}).
Thus we suggest that the comparison of size and surface brightness of \oiii regions with observations can help constrain the physical models of radiative feedback in high-$z$ galaxies. 
Note that, however, if the metal distribution is still confined to only star-forming regions, it is likely to be difficult to distinguish the models.
Therefore, the comparison test should be applied to massive metal-rich galaxies.

At $z<10$, $\langle f_{\rm esc} \rangle$ decreases to $\sim 1\,\%$ as the halo mass increases, which is consistent with the literature \citep[e.g.][]{Yajima11,Yajima14, Paardekooper15}.
In addition, as shown in A19, dusty gas is held in the massive halo against SN feedback, and the dust efficiently absorbs ionizing photons.

The bottom panel of Fig.\,\ref{fig:halo11} shows the \oiii and \cii  luminosities.
The \oiii line is emitted only in the star-bursting phase because O$^{2+}$ ions exist in {\sc H\,ii} regions formed by massive stars.
We find that $L_{\rm [O_{\III}]}$ fluctuates with intermittent star formation in the range of $10^{40}-10^{42}\,{\rm erg~s^{-1}}$ at $z<10$ for Halo-11.
Meanwhile, the main source of \cii emission is neutral gas, thus it is continuously emitted even in the outflowing phase.
The contribution of \cii emission from {\sc H\,ii} regions to the total luminosity (dotted line) increases during the  star-bursting phase, but it does not exceed 50\,\% over all redshifts. 
It gradually decreases with time due to dust absorption and results in $\lesssim 2\,\%$ at $z=6$ .

We note that in some snapshots Halo-11 has high \oiii luminosity at very high redshifts ($z>10$) even when the escape fraction is $\sim 0.4$. This is because the stars have low-metallicity ($Z \lesssim 0.01\,\Zsun$) and emit ionizing photons efficiently. Thus one needs to consider stellar populations carefully when discussing the $L_{\rm [O_{\III}]}$--SFR relation (see next sub-section).

\subsection{Metal-line luminosity vs. SFR}
\label{sec:lum_SFR}

\begin{figure}
\begin{center}
\includegraphics[width=1.07\columnwidth]{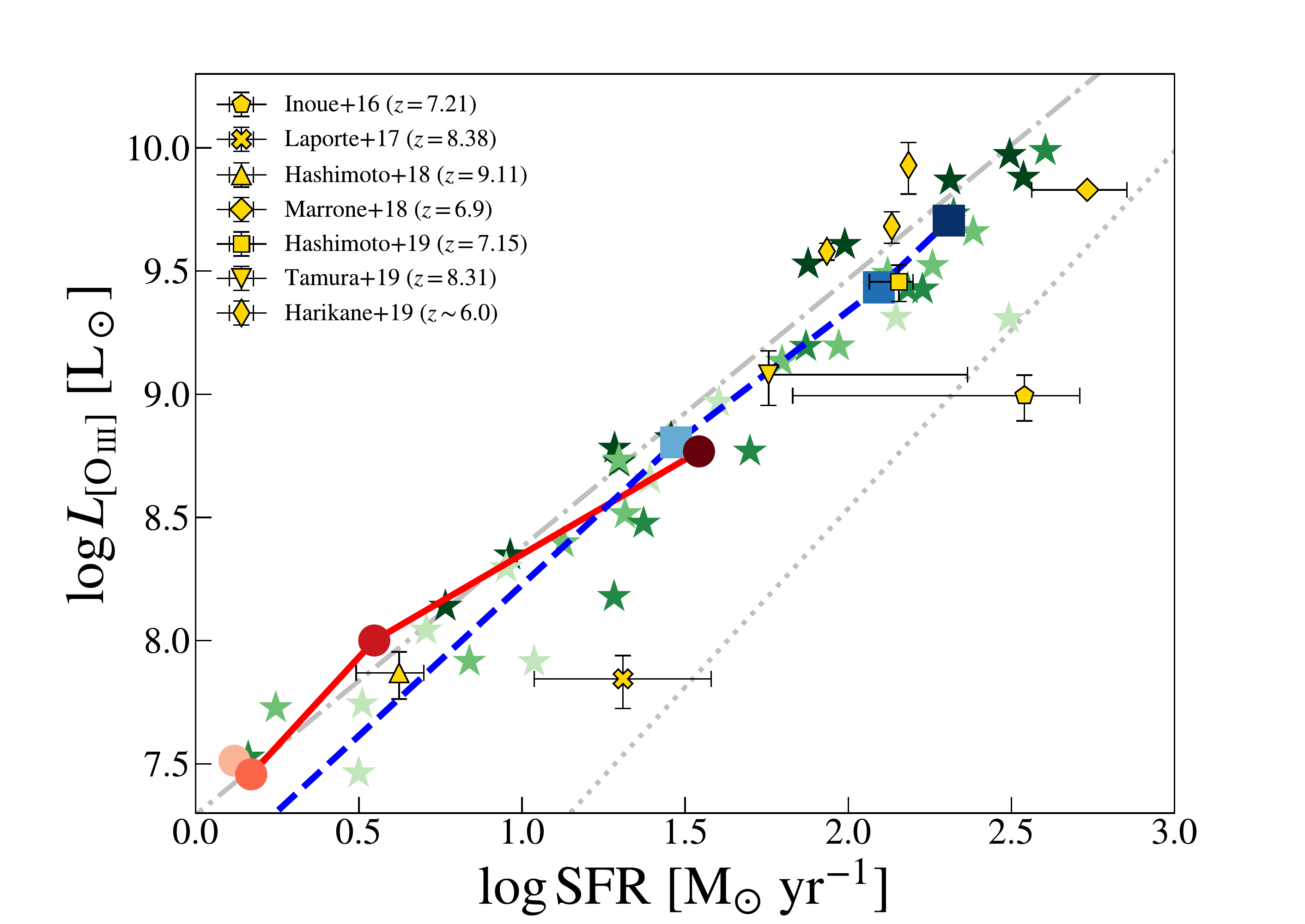}
\caption{Relation between SFR and \oiii luminosity. Red circles (solid line) represent the evolution for Halo-11 at $z=9,~8,~7,~6$ (from lighter to darker).
Blue squares (dashed line) are for Halo-12. Green stars are for other 10 samples at $z=6-9$. Yellow symbols represent the observed high-$z$ galaxies. Gray dotted line and dot-dashed line are for the local starburst galaxies and metal-poor dwarf galaxies, respectively \citep{DeLooze14}.
}
\label{fig:oiii}
\end{center}
\end{figure}

\begin{figure}
\begin{center}
\includegraphics[width=1.07\columnwidth]{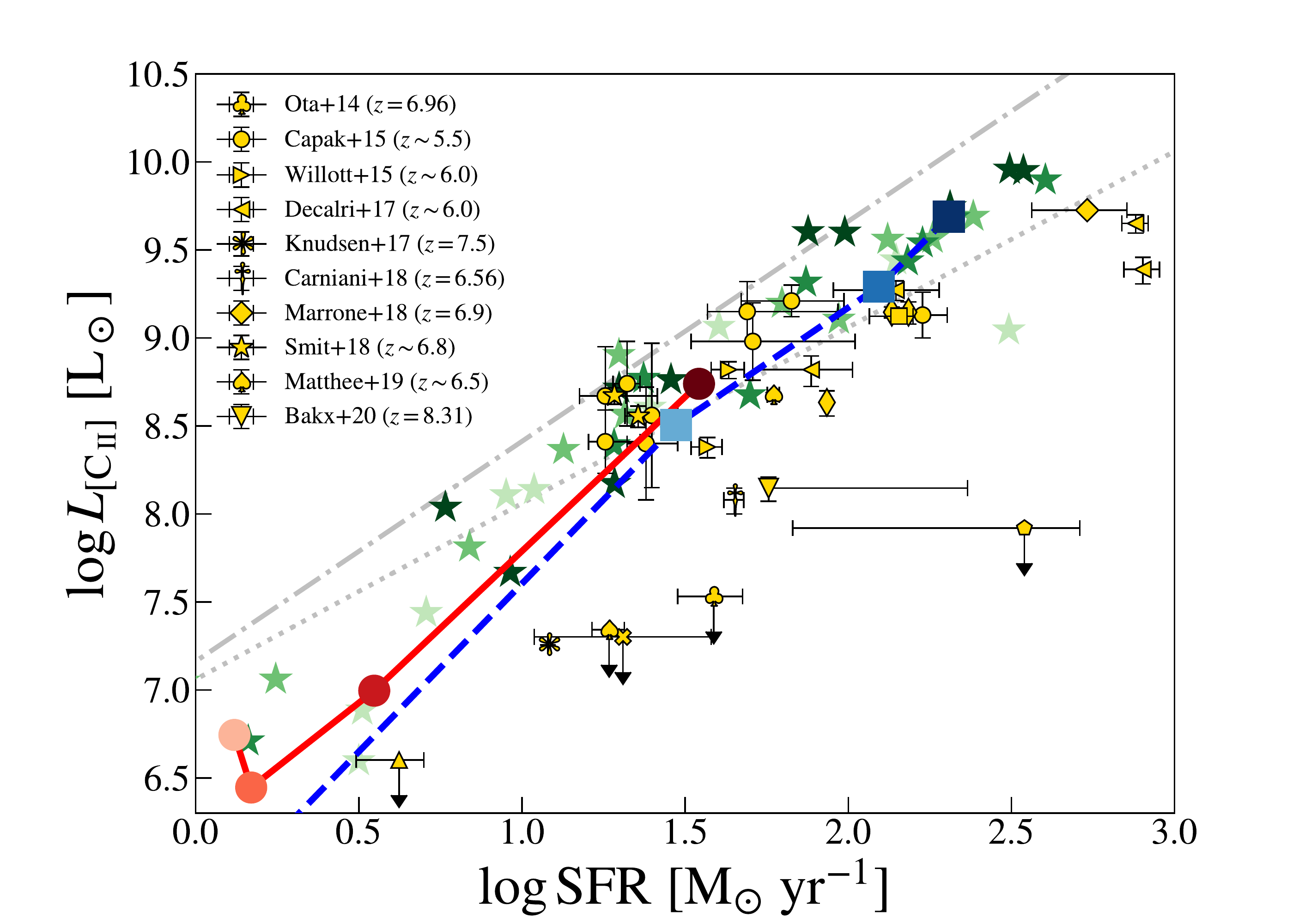}
\caption{
Relation between SFR and \cii luminosity. The meaning of symbols is the same as in Fig.\,\ref{fig:oiii}. The symbols of \citet{Decarli17} represent companion galaxies of quasars. Some samples \citep{Ota14,Carniani18,Matthee19} are not detected in IR. To estimate their total SFR, we simply use the upper limit of ${\rm SFR_{\rm IR}}$.
}
\label{fig:cii}
\end{center}
\end{figure}

Figures\,\ref{fig:oiii} and \ref{fig:cii} present \oiii and \cii luminosities vs. SFR for Halo-11 and Halo-12 at $z=6,~7,~8$ and 9.
Our simulations reproduce the observed luminosities well at $z \gtrsim 6$.
In particular, Halo-12 shows $L_{\rm [O_{\III}]}=2.3\times 10^{9}\,{\rm \Lsun}$, $L_{\rm [C_{\II}]}=1.7\times 10^{9}\,{\rm \Lsun}$, and $L_{\rm IR}=6.9\times 10^{11}\,{\rm \Lsun}$ at $z=7$, which are remarkably consistent with B14-65666 at $z=7.15$ ($L_{\rm [O_{\III}]}=2.9\times 10^{9}\,{\rm \Lsun}$, $L_{\rm [C_{\II}]}=1.3\times 10^{9}\,{\rm \Lsun}$, and $L_{\rm IR}=6.2\times 10^{11}\,{\rm \Lsun}$) as reported by \citet{Hashimoto19a}.
\oiii luminosities in our simulations cannot reproduce the
galaxies with lower luminosities observed in \citet{Inoue16} and \citet{Laporte17}.
Also our simulations have higher \cii luminosities than some observed galaxies with ${\rm log~SFR} < 1.5$. The origin of this deviation requires simulating a larger sample of galaxies which we aim in future work.

Also we compare our results with the relation of local galaxies derived in \citet{DeLooze14}.
They presented calibrations of the SFR--$L_{\rm line}$ relations in the range of $-3 \lesssim {\rm \log{\rm SFR\,[\Msunyr]}} \lesssim 2$ using low-metal galaxies from the {\it Herschel} Dwarf Galaxy Survey and other local samples from previous FIR line measurements \citep{Brauher08, Parkin13, Sargsyan12, Diaz-Santos13, Farrah13, Gracia-Carpio11}.
The \oiii luminosities in our simulations are similar to those of metal-poor dwarf galaxies rather than local starburst galaxies. 
This result is consistent with \citet{Moriwaki18}, but is opposite to those of \citep{Olsen17,Katz19}.
The chemical abundance pattern of galaxies at $z>6$ is dominated by Type-II SNe, and quite different from the solar neighbourhood.
In the calculation of \oiii luminosity, we use the oxygen abundance of each gas particle, which is the same treatment as in \citet{Moriwaki18} but not as in \citet{Olsen17} and \citet{Katz19}.
\citet{Katz19} showed that their SFR--$L_{\rm line}$ relations had a good agreement with the local galaxies of \citet{DeLooze14}, however their simulations did not track element abundances separately.
The oxygen abundance (mass-weighted mean) of Halo-12 evolves from $12+\log{\rm (O/H)}= 6.6$ at $z=9$ to $8.9$ at $z=6$, and the range is close to that of local relation of $12+\log{\rm (O/H)}= 7.14 - 8.43$ \citep{DeLooze14}.
Thus we suggest that the physical state of the ISM in the first galaxies might be similar to that of local dwarf galaxies except for massive high-$z$ galaxies like Halo-12 at $z \sim 6$.

From the least-square fitting to all samples (Halo-11, Halo-12, MHaloes, LHaloes) at $z=6-9$, 
we derive the following relations: 
\begin{equation}
\label{eq:sfr_oiii}
    \log{(L_{\rm [O_{\III}]}\,{\rm [\Lsun]})} = 
    7.23 + 1.04 \log{\rm (SFR\,[\Msun~yr^{-1}])},
\end{equation}
\begin{equation}
\label{eq:sfr_cii}
    \log{(L_{\rm [C_{\II}]}\,{\rm [\Lsun]})} = 
    6.38 + 1.47 \log{\rm (SFR\,[\Msun~yr^{-1}])}.
\end{equation}
We find that \oiii luminosity is linearly proportional to the SFR, because most of the ionizing photons are absorbed by gas ($f_{\rm esc}^{\,\rm ion}\lesssim 0.1$) for all of the galaxies and the volume of {\sc H\,ii} regions (\oiii emitting regions) linearly increases. 
These fits nicely match the observed results at $z \sim 6-9$ \citep{Harikane19}.
The $L_{\rm [O_{\III}]}$--SFR relation at high-$z$ is very close to the relation for local metal-poor dwarf galaxies,
while the slope of $L_{\rm [C_{\II}]}$--SFR relation is steeper than the local ones \citep[1.0 for starburst galaxies and 1.25 for metal-poor galaxies,][]{DeLooze14}.
Therefore, this indicates that neutral gas distribution in high-redshift galaxies is different from local ones.
In addition, these results are also supported by a more statistical study using semi-analytical models \citep{Lagache18}.

Note that if  efficient absorption is a common feature for high-$z$ galaxies, any deviation from a Chabrier IMF would be  reflected in the first term of Eq.\,(\ref{eq:sfr_oiii}). Noting that $L_{\rm [O_{\III}]} \propto \dot{N}_{\rm{ion}} \propto \rm{SFR}$.
In other words, a top-heavy IMF predicts higher \oiii luminosity at a given SFR, because more gas clouds would be ionized.
Therefore a good observational estimate of this relationship can give us insights  about the physics of star formation in high-$z$ galaxies. 
The same is also true for Eq.\,(\ref{eq:sfr_cii}), which includes the contribution from accreting cold gas ($\sim 10^{4}\,{\rm K}$) \citep[e.g.][]{Dekel06}. 
A higher normalization of Eq.\,(\ref{eq:sfr_cii}) points toward higher accretion rates from cold accretion, and we can also expect this relation to change with redshift slightly as accretion rates are higher at earlier times.
Additionally, a top-heavy IMF can also change the gas distribution significantly via stronger feedback. If a strong gas outflow is induced, ionizing photon can escape efficiently, resulting in lower \oiii luminosity. 
Therefore, further simulations are needed to provide stronger constraints  on the impact of the IMF.

\begin{figure}
\begin{center}
\includegraphics[width=\columnwidth]{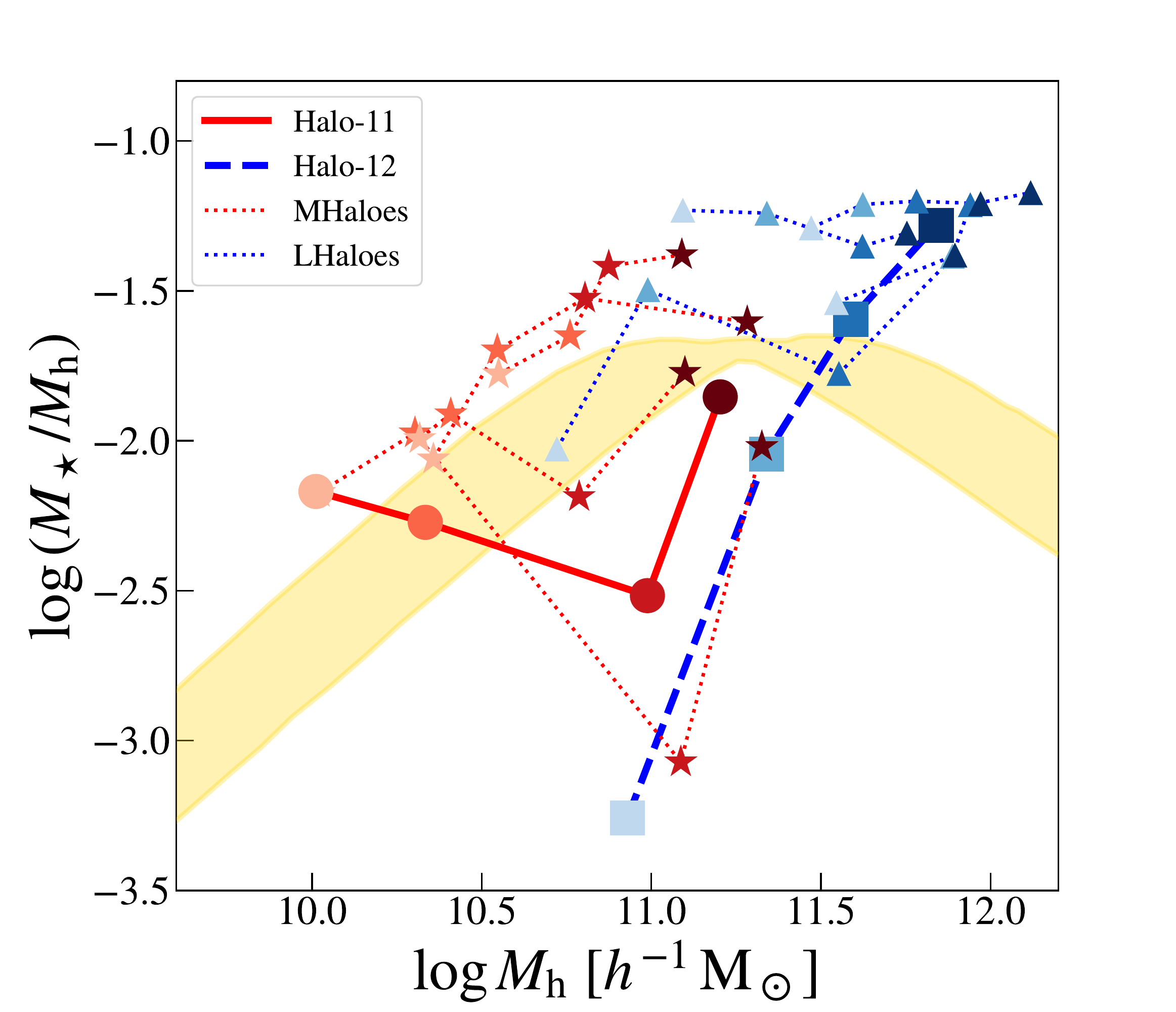}
\caption{
Stellar-to-halo-mass ratio (SHMR) as a function of halo mass.
The meanings of red circles (Halo-11) and blue squares (Halo-12) are the same as in Fig.\,\ref{fig:oiii}.
The red stars and blue triangles represent MHaloes and LHaloes at $z=9,~8,~7,~6$ (from lighter to darker color).
The yellow shade shows the observational fits at $z=6-8$ based on the abundance matching method \citep{Behroozi13}.
}
\label{fig:shmr}
\end{center}
\end{figure}

Finally, we compare the stellar masses of different halos at $z=6-9$ to examine the effect of different star formation histories. 
Figure~\ref{fig:shmr} shows the stellar-to-halo-mass ratio (SHMR) as a function of halo mass.
Halo-11 and MHaloes roughly match with the observational SHMR at $z=6-8$ \citep[the so-called abundance matching result by][]{Behroozi13}.
However, Halo-12 and LHaloes have slightly higher SHMR than the Behroozi result. 
One of the interpretations of this is that we need more effective feedback processes, e.g., stellar radiation or AGN feedback, which suppresses star formation in massive haloes.
We also note that the observations might be highly biased due to lack of large high-$z$ samples.

\subsection{\oiii and \cii luminosity ratio}
\label{sec:lum_ratio}

In this subsection, we focus on the \oiii and \cii luminosity ratio. 
\citet{Hashimoto19a} showed the negative correlation between $L_{\rm [O_{\III}]} / L_{\rm [C_{\II}]}$ and bolometric luminosity $L_{\rm bol}$ for galaxies at $z>6$, in range of $10^{10}\,\Lsun < L_{\rm bol} < 10^{14}\,\Lsun$ \citep[see also][]{Marrone18}.
The luminosity ratio can reflect the multi-phase ISM structure \citep[e.g.][]{Cormier12}.
\citet{Cormier15} found the same trend in local star-forming galaxies, and suggested that it depended on gas metallicity.
Here we suggest that the origin of the negative correlation of high-$z$ galaxies is the carbon enrichment by AGB stars and the decrease of {\sc H\,ii}  fraction.

\begin{figure}
\begin{center}
\includegraphics[width=\columnwidth]{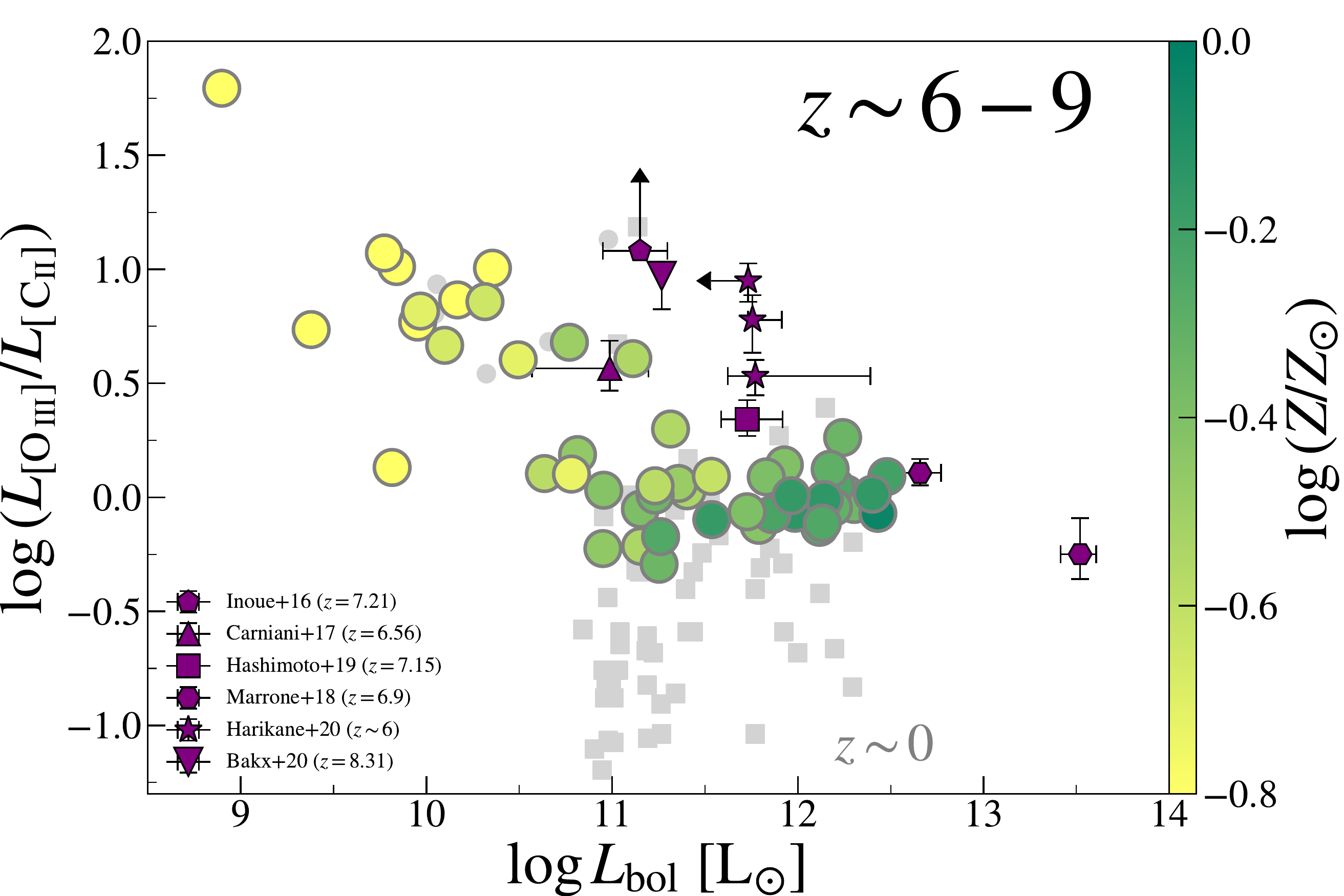}
\caption{Relation between $L_{\rm [O_{\,\III}]}/L_{\rm [C_{\,\II}]}$ ratio and bolometric luminosity ($L_{\rm bol}=L_{\rm UV}+L_{\rm IR}$). Filled circles represent our simulation samples at $z\approx 6-9$. The color is scaled by gas metallicity, which is an indicator of galactic evolution. The origin of the anti-correlation is carbon enrichment by AGB stars (see the main text). The purple symbols represent observed galaxies at $z\sim 6-9$ \citep{Inoue16,Carniani17,Marrone18,Hashimoto19a,Harikane20,Bakx20}. The grey symbols are for 
the local galaxies \citep{Howell10,Madden13,DeLooze14,Cormier15,Diaz-Santos17}.
}
\label{fig:ratio_all}
\end{center}
\end{figure}

\begin{figure}
\begin{center}
\includegraphics[width=\columnwidth]{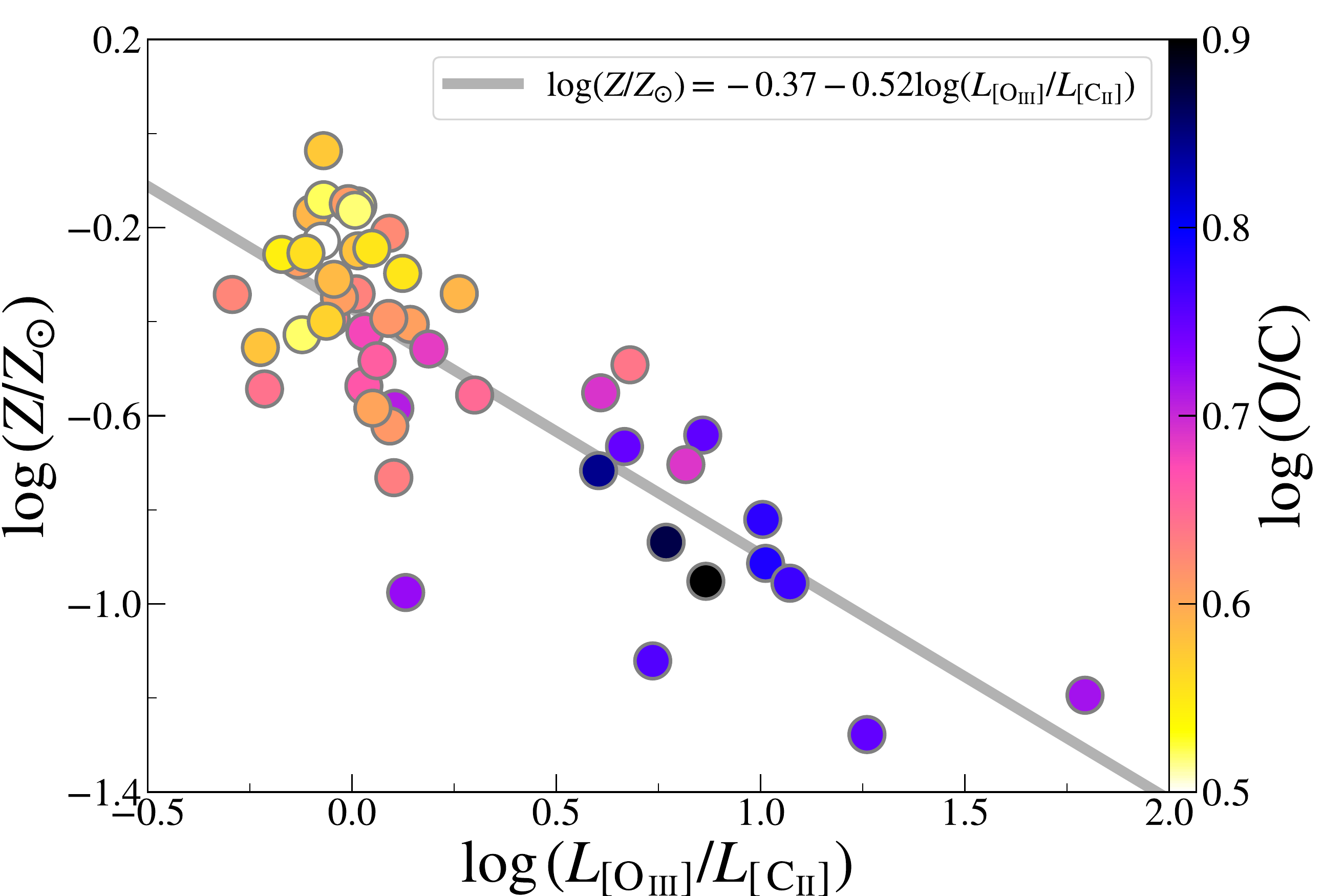}
\caption{Relation between $L_{\rm [O_{\,\III}]}/L_{\rm [C_{\,\II}]}$ ratio and total metallicity in our simulation samples at $z=6-9$. The color is scaled by O/C abundance ratio.
Gray line shows result of the least-square fitting.
}
\label{fig:ratio_vs_metal}
\end{center}
\end{figure}

Figure~\ref{fig:ratio_all} presents the $(L_{\rm [O_{\III}]} / L_{\rm [C_{\II}]})$--$L_{\rm bol}$ relation at $z = 6-9$, where $L_{\rm bol}$ is measured from $L_{\rm UV}+L_{\rm IR}$. 
We find that $L_{\rm [O_{\III}]} / L_{\rm [C_{\II}]}$ ratio decreases from $\sim 10$ to $\sim 1$ in the range of $10 \lesssim \log{\,(L_{\rm bol}\,[\Lsun])} \lesssim 12$ with increasing metallicity from $\sim 0.1\,\Zsun$ to $\sim 1\Zsun$.
The $\log{(L_{\rm [C_{\II}]}/{\rm SFR}\, {\rm [\Lsun/\Msunyr]})}$ increases from $6.1$ to $7.8$, 
while $\log{(L_{\rm [O_{\III}]}/{\rm SFR})}$ is constant ($\sim 7.25 \pm 0.25$).
The lower-mass galaxies has higher O/C ratio because Type-II SNe efficiently enrich the ISM with oxygen than carbon.
As the galaxy evolves, low- and intermediate-mass stars begin to eject carbon-rich winds by dredge-up events of AGB phases \citep[see also][]{Berg19}.
The $\log{\rm (O/C)}$ decreases from $\sim  0.9$ to $\sim  0.5$ when metallicities increase from $\sim 0.1\,\Zsun$ to $\sim 1\Zsun$, 
therefore the $L_{\rm [O_{\III}]} / L_{\rm [C_{\II}]}$ ratio decreases  (Figure~\ref{fig:ratio_vs_metal}).
To confirm the validity of this scenario,
we use the Chemical Evolution Library \citep[CELib,][]{Saitoh17}, and compute $\log{\rm (O/C)}$ in a star-forming cloud with $Z=0.1\,\Zsun$ and the Chabrier IMF. 
The $\log{\rm (O/C)}$ decreases from $\sim 1.0$ to $\sim 0.5$ 
when the age exceeds $\sim 0.5\,{\rm Gyr}$ which corresponds to half of the cosmic time at $z\sim 6$.
Of course, in the actual hydrodynamic simulation, we have inflow of low-metallicity gas as well as metal-enriched outflows, so the situation is not as simple as the CELib result. However, the line luminosities are weighted by the densities of each element, and more dominated by the higher density gas near the galactic center, which is less affected by the outer gas. At least, the qualitatively trend of CELib result is consistent with our simulation result.

Our simulations are in good agreement with high-$z$ observations. 
Also, $L_{\rm [O_{\III}]} / L_{\rm [C_{\II}]}$ ratios decreases as the metallicity increases, of which the trend is similar to local dwarf galaxies \citep{Cormier15}. However, the absolute values of the ratios are higher than the local galaxies (grey symbols).
Our samples have high ionization parameter ($-2.2\lesssim \log{U} \lesssim -1.5$, where $U$ is mean ionization parameter weighted by \oiii luminosity) and high volume fraction of {\sc H\,ii} regions ($0.64\lesssim f_{\rm H_{\,\II}} \lesssim 1.0$).
Using {\sc Cloudy} models, \citet{Harikane19} suggested that galaxies with high-$U$ and low covering fraction of PDR can explain higher $L_{\rm [O_{\III}]} / L_{\rm [C_{\II}]}$ ratio than local galaxies.
Our results are consistent with Harikane's picture of the first galaxies.

\begin{figure}
\begin{center}
\includegraphics[width=\columnwidth]{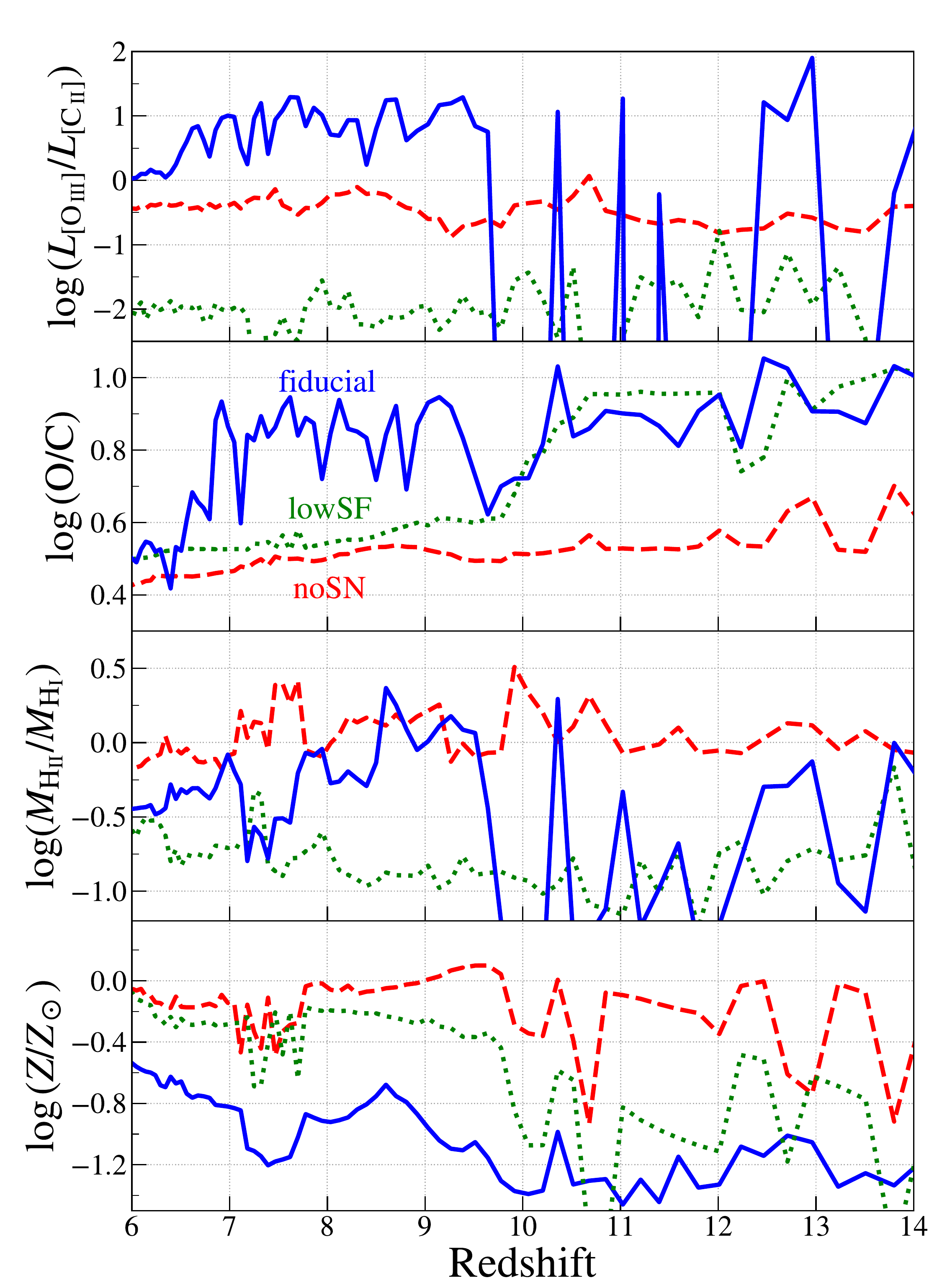}
\caption{Redshift evolution of $L_{\rm [O_{\,\III}]}/L_{\rm [C_{\,\II}]}$ ratio ({\it top}), density-weighted mean O/C abundance ratio ({\it second}), mass ratio of {\sc H\,ii} to {\sc H\,i} regions ({\it third}), and gas metallicity within the halo ({\it bottom}). The solid blue line represents evolution of Halo-11, and red dashed and green dotted lines are for Halo-11-noSN and Halo-11-lowSF cases, respectively. 
}
\label{fig:ratio_halo11}
\end{center}
\end{figure}

Figure\,\ref{fig:ratio_halo11} describes more details of redshift evolution of $L_{\rm [O_{\III}]} / L_{\rm [C_{\II}]}$ ratio.
The top panel shows that $L_{\rm [O_{\III}]} / L_{\rm [C_{\II}]}$ ratio largely fluctuates with intermittent star formation in case of Halo-11 (blue line).
But it gradually decreases with decreasing redshift due to carbon enrichment by AGB winds (second panel).
Due to continuous star formation at $z<10$, gas in Halo-11 is highly ionized ($f_{\rm H_{\II}}\gtrsim 0.7$).
However, the gas density in {\sc H\,ii} regions is high, because the massive haloes keep gas against SN feedback, resulting in lower-$U$.
The third panel shows the mass ratio of {\sc H\,ii} to {\sc H\,i} regions. At lower redshifts, the dense gas rapidly recombines and reduces the mass ratio, resulting in decreasing the luminosity ratio because the dominant \oiii and \cii sources are {\sc H\,ii} and {\sc H\,i} regions, respectively.
The bottom panel shows that the gas metallicity in the halo gradually  increases from $z\sim 10$ towards lower redshift, because metal-enriched gas ejected by SN feedback can fall back to the halo due to the deeper gravitational potential well. 

Note that from the equation of line emission, the $L_{\rm [O_{\III}]} / L_{\rm [C_{\II}]}$ ratio depends on the O/C abundance ratio, mass ratio of $M_{\rm H\,II}/M_{\rm H\,I}$ and a factor of the Einstein's $C$-coefficient.
At $z=6$, both of the O/C ratio and $M_{\rm H\,II}/M_{\rm H\,I}$ ratio of Halo-11 become $\sim 0.4$ times as much as at $z=7$, resulting in largely decreasing the $L_{\rm [O_{\III}]} / L_{\rm [C_{\II}]}$ ratio. The evolution of density and temperature also contributes to this decline.

We also compare the $L_{\rm [O_{\III}]} / L_{\rm [C_{\II}]}$ ratio with that of lowSF (green) and noSN (red) cases.
In the lowSF case, we set a lower value for the coefficient of Kennicutt--Schmidt relation ($A$) than that of the fiducial case, which induces higher gas density (see also, Y17).
The SFR becomes similar in both cases by feedback processes, but the {\sc H\,ii} volume fraction in the lowSF case is lower, because the dense gas rapidly recombines (see Sec.\,\ref{sec:model}).
The high density also makes the energy injection by SNe inefficient due to radiative cooling, thus the metal enrichment proceeds faster.
These factors lead to very low $L_{\rm [O_{\III}]} / L_{\rm [C_{\II}]}$ ($\sim -2.0$ at $z<10$).
On the other hand, the noSN case has similar gas densities as the fiducial case, but
SN feedback does not evacuate gas from the halo, resulting in rapid metal enrichment. 
The O/C abundance ratio stays at low values, resulting in low $L_{\rm [O_{\III}]} / L_{\rm [C_{\II}]}$  ($\sim -0.5$).

Metallicity in high-$z$ galaxies are observationally estimated  using the attenuation by Fe ions in the UV part of the stellar spectrum.
However, \citet{Cullen17} shows that simulated galaxies at $z\gtrsim 5$ are $\alpha$-element enhanced, which is consistent with the observational study of \citet{Steidel16}.
They found a factor of $\sim 5$ difference between stellar and nebular metallicities.
Thus the observational metallicity estimate based on Fe abundance does not trace the total metallicity in high-$z$ galaxies.
Figure~\ref{fig:ratio_vs_metal} shows that observations of $L_{\rm [O_{\,\III}]}/L_{\rm [C_{\,\II}]}$ ratio could be a useful tool to estimate the total metallicity of galaxies at high-$z$ more precisely.
Using a least-square fitting for $z=6-9$ galaxies with $-0.5<\log{(L_{\rm [O_{\,\III}]}/L_{\rm [C_{\,\II}]})} < 2.0$, we derive the following relation for the total metallicity: 
\begin{equation}
    \log{(Z/\Zsun)} = -0.37 - 0.52 \log{(L_{\rm [O_{\,\III}]}/L_{\rm [C_{\,\II}]})}.
\end{equation}
For example, the metallicity of B14-65666 at $z=7.15$ \citep{Hashimoto19a} is estimated to be  $Z=0.29\Zsun$ using the above relation.
Note that the dispersion is large with $\gtrsim 0.2$\,dex, and there are some outliers. This can be due to other factors, e.g., ionization degree, or the small sample sizes. We will investigate in more detail statistically in the future work.

\subsection{The ($L_{\rm line} / L_{\rm IR}$)--$L_{\rm IR}$ relations}
\label{sec:line_ir}

In addition to the ionization of hydrogen and metals, UV radiation from young stars also heats up dust, resulting in IR thermal emission. Therefore, both \oiii and IR luminosities are likely to be related with the SFR. However, depending on the ratio of absorbed energy of ionizing photons between gas and dust, the luminosity ratio changes. 
Local observations show that $L_{\rm [O_{\III}]} / L_{\rm IR}$ and $L_{\rm [C_{\II}]} / L_{\rm IR}$ decrease as $L_{\rm IR}$ increases
\citep{DeLooze14,Cormier15,Diaz-Santos14}.
\citet{Cormier15} point out that extended dwarf galaxies have higher $L_{\rm [O_{\III}]} / L_{\rm IR}$ and suggest the large volume fraction of ionized regions be the reason for this.
In addition, recent observations show a similar trend even for high-$z$ galaxies \citep[e.g.][]{Tamura19}.
Here we study the origin of the correlations in $(L_{\rm line} / L_{\rm IR})$--$L_{\rm IR}$ relation.

\begin{figure}
\begin{center}
\includegraphics[width=\columnwidth]{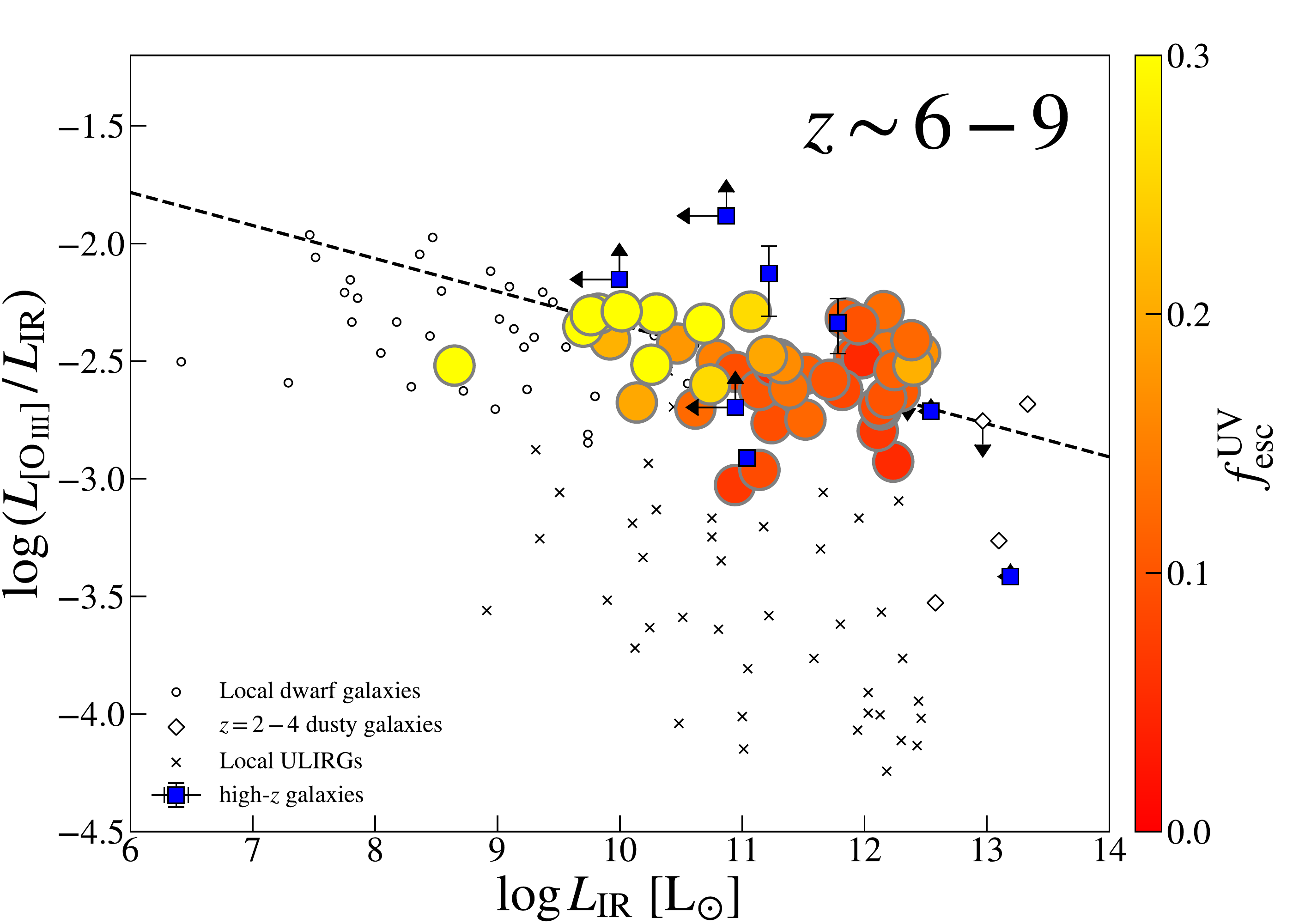}
\caption{Relation between luminosity ratio $L_{\rm [O_{\III}]}/L_{\rm IR}$ and total IR luminosity $L_{\rm IR}$. Filled circles represent our simulation samples at $z=6-9$. The color is scaled by escape fraction of UV photons.
The black dashed line is derived from the fitting functions given in Eq.\,\ref{eq:sfr_oiii} and \ref{eq:sfr_ir}.
Open small circles represent local dwarf galaxies \citep{Madden13,Cormier15}. 
Open diamonds are for dusty star-forming galaxies at $z\sim 2-4$ \citep{Ferkinhoff10,Ivison10,Valtchanov11,Vishwas18}.
Small crosses are for from local spiral galaxies to ULIRGs \citep{Herrera-Camus18a,Herrera-Camus18b}.
Blue squares are for $z\gtrsim 6$ galaxies  \citep{Inoue16,Carniani17,Laporte17,Hashimoto18a,Marrone18,Tamura19,Hashimoto19a}.}
\label{fig:oiii_ratio}
\end{center}
\end{figure}

Figure~\ref{fig:oiii_ratio} shows the relation between $L_{\rm [O_{\III}]}/L_{\rm IR}$ and $L_{\rm IR}$ at $z=6-9$.
There is a weak negative correlation that is consistent with local star-forming galaxies, $z\sim 2-4$ dusty star-forming galaxies and ultra-luminous infrared galaxies (ULIRGs).
At $z<10$, most of the ionizing photons are absorbed by the gas even in less massive galaxies, resulting in a low escape fraction ($f_{\rm esc}^{\rm ion} \lesssim 0.1$, see Fig.\,\ref{fig:halo11}) and a linear $L_{\rm [O_{\III}]}$--SFR relation (equation\,\ref{eq:sfr_oiii}). Meanwhile, UV continuum photons can escape through the direction of low dust column density in the outflowing phase ($\tau_{\rm UV} \lesssim 1$).
We find that the negative correlation is closely related to the escape fraction of UV photons.
In A19, using all of the galaxies including main and satellites in the zoom-in boxes of Halo-11 and Halo-12, we presented the $L_{\rm IR}$--SFR relation at $z\sim 7$ (Equation\,6 in A19):
\begin{equation}
\label{eq:sfr_ir}
    \log{(L_{\rm IR}\,{\rm [\Lsun]})} = 
    9.5 + 1.21 \log{\rm (SFR\,[\Msun~yr^{-1}])}.
\end{equation}
If the UV light is completely reprocessed into the IR-band, the slope must be unity because UV luminosity is linearly proportional to SFR \citep[e.g.,][]{Kennicutt98}. However, the high UV escape fraction of low-mass galaxies reduces the IR luminosity, resulting in as steeper slope.
Combining Eq.\,(\ref{eq:sfr_oiii}) and \,(\ref{eq:sfr_ir}), we plot the thick yellow band on the figure.
It roughly reproduces our simulation results and observations of high-$z$ galaxies, while some high-redshift galaxies are distributed far away from our model results.
In addition, our galaxies have high volume fraction of {\sc H\,ii} regions ($0.7 < f_{\rm H_{\,\II}} < 1$), which is responsible for higher $L_{\rm [O_{\III}]}/L_{\rm IR}$ than in local star-forming galaxies at a specific $L_{\rm IR}$.

\begin{figure}
\begin{center}
\includegraphics[width=\columnwidth]{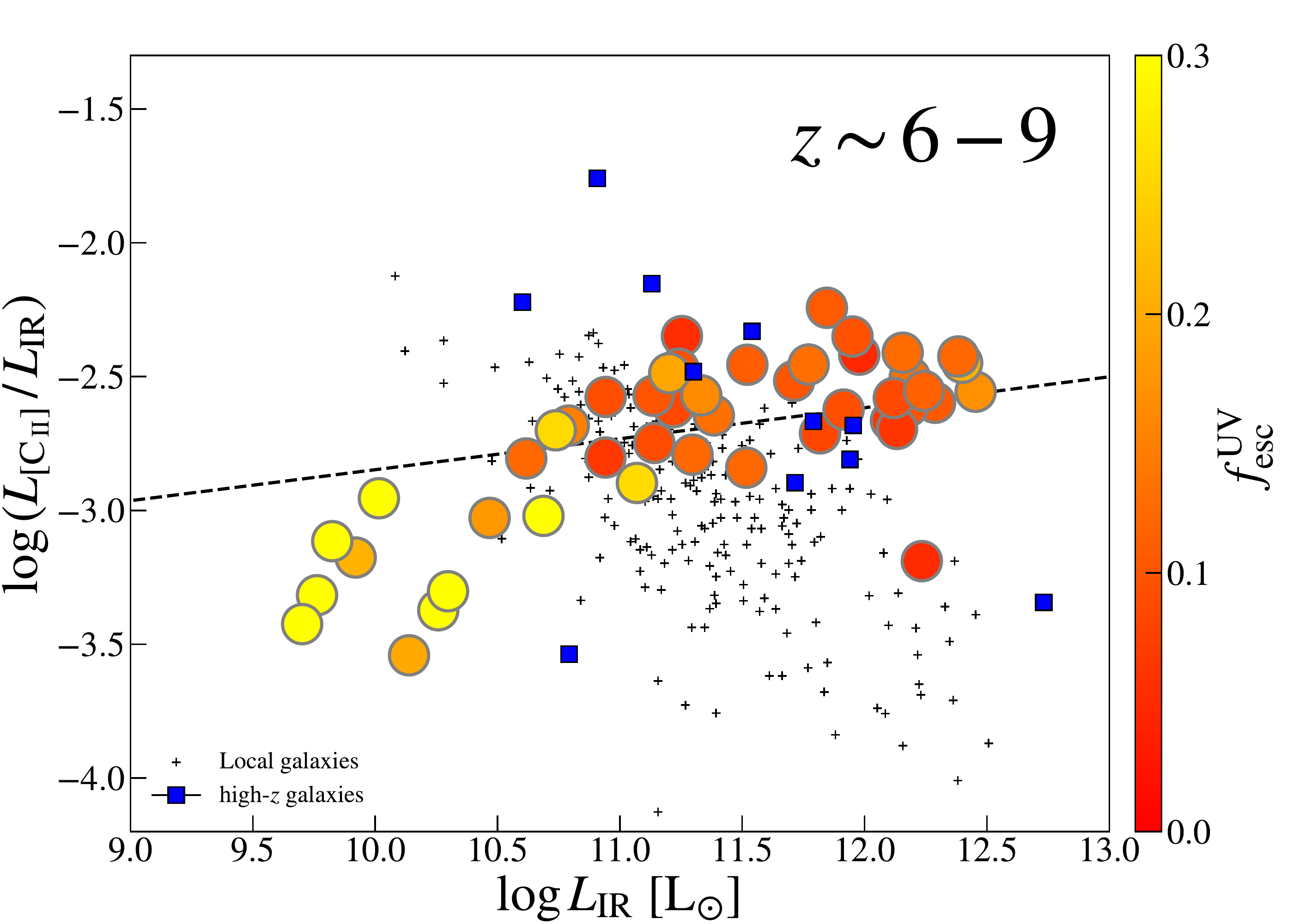}
\caption{Relation between $L_{\rm [C_{\II}]}/L_{\rm IR}$ and $L_{\rm IR}$. The meaning of filled circles is the same as in Fig.\,\ref{fig:oiii_ratio}.
The black dashed line is derived from the fitting functions given in Eq.\,\ref{eq:sfr_cii} and \ref{eq:sfr_ir}.
 Small plus symbols represent the local galaxies from LIRGs to ULIRGs \citep{Diaz-Santos14}. Blue squares are for $z\gtrsim 5$ galaxies \citep{Capak15,Willott15,Knudsen17,Decarli17,Hashimoto19a}.}
\label{fig:cii_ratio}
\end{center}
\end{figure}

Figure\,\ref{fig:cii_ratio} shows ($L_{\rm [C_{\II}]} / L_{\rm IR}$)--$L_{\rm IR}$ relation.
The model based on the combination of Eq.\,(\ref{eq:sfr_cii}) and (\ref{eq:sfr_ir}) is plotted as a thick yellow band.
It increases with $L_{\rm IR}$ because of the steeper slope of the $L_{\rm [C_{\II}]}$--SFR relation than that of  the $L_{\rm IR}$--SFR relation. Our simulations match the model nicely. 
However, the luminosity ratios of observed galaxies decreases as $L_{\rm IR}$ unlike in our simulations. 
This discrepancy can be explained by the dust effect. 
\citet{Luhman03} suggested the `dust-bounded model', where the dust in {\sc H\,ii} regions efficiently absorbs  UV photons, resulting in inefficient photo-electric heating of polycyclic aromatic hydrocarbons (PAHs) in PDRs.
Therefore, the thermally-balanced \cii cooling also becomes inefficient.
Also, if the dust is positively charged, the photo-electric heating and \cii cooling can be suppressed \citep{Wolfire90,Luhman03}. 
Combining \cii 158$\,{\rm \mu m}$ and CO (3-2) observations at $z\sim 3$, \citet{Rybak19} suggests that high-temperature saturation of C$^{+}$ level populations due to a strong UV radiation field induce the \cii deficit \citep{Munoz16}.
Our simulations do not include  photo-electric heating. Given that the photo-electric heating is included, the \cii luminosities of bright galaxies could be decreased because a part of far-UV photons are absorbed in {\sc H\,ii} regions near the star, and the temperature of {\sc H\,i} regions becomes lower. Also, \cii luminosities of faint galaxies could be increased by the photo-electric heating because in their dust-less ISM most of FUV photons reach to {\sc H\,i} regions and increase the temperature, inducing the \cii emission. These effects can change the trend. 
In addition, we assume that all carbons in neutral hydrogen gas cells are in the C$^{+}$ state. In high-density low-temperature regions, a part of carbons form CO molecules. If its effect is not negligible, we overestimate the \cii luminosity.

We also note that the observed IR luminosity of high-$z$ galaxies are not constrained well. 
Since most galaxies were detected with a single sub-mm band, the dust mass and total IR luminosity were estimated with assumed dust temperatures \citep[e.g.][]{Watson15}.
In our previous paper, we showed that dust temperature of high-$z$ galaxies could be higher than that of local ones \citep[see also,][]{Ma19}.
Currently, the multi-wavelength observations of dust continuum have advanced.
\citet{Bakx20} observed a $z=8.31$ galaxy \citep[MACS0416\_Y1,][]{Tamura19} in three sub-mm wavelength, and showed that the dust temperature should be warm ($\gtrsim 80\,{\rm K}$).
Meanwhile, \citet{Harikane19} showed that their two galaxies at $z\sim 6$ have similar dust temperature with the local ones ($\sim 30\,{\rm K}$.)
Future observations will reveal dust temperature of high-$z$ galaxies, allowing us to estimate the IR luminosity more accurately.

\subsection{Galaxy size measured by \cii line}
\label{sec:dist}

The size of galaxy was classically considered to be determined by the conservation of angular momentum of accreting gas \citep{Mo98}.
However, recent simulations showed that stellar feedback re-distributes angular momentum and changes galaxy sizes \citep[][Y17]{Genel15}.　
Therefore the galaxy morphology can be used as a test of theoretical feedback models.
Recently, \citet{Fujimoto19} stacked the ALMA data of 18 galaxies at $z\sim 5-7$, and detected the \cii emission extended over $\sim 10\,{\rm kpc}$. 
The effective radius was larger than the disk scale measured from the rest-UV and FIR continuum emissions. 

Our simulated galaxies have disk structures at $z\lesssim 10$, and the sizes are affected by SN feedback as described in Y17.
In A19, we showed that the UV half-light radius changed within the range of $\sim 1-10\,\%$ of virial radius $(r_{\rm vir})$ due to intermittent star formation history.
The UV radial profile was dominated by the central star-forming regions in star-burst phases.
On the other hand, in outflowing phases, the gas is ejected from the centre and the gravitational potential becomes shallower, resulting in an extended distribution of residual high-mass stars and UV radial profile.

\begin{figure}
\begin{center}
\includegraphics[width=\columnwidth]{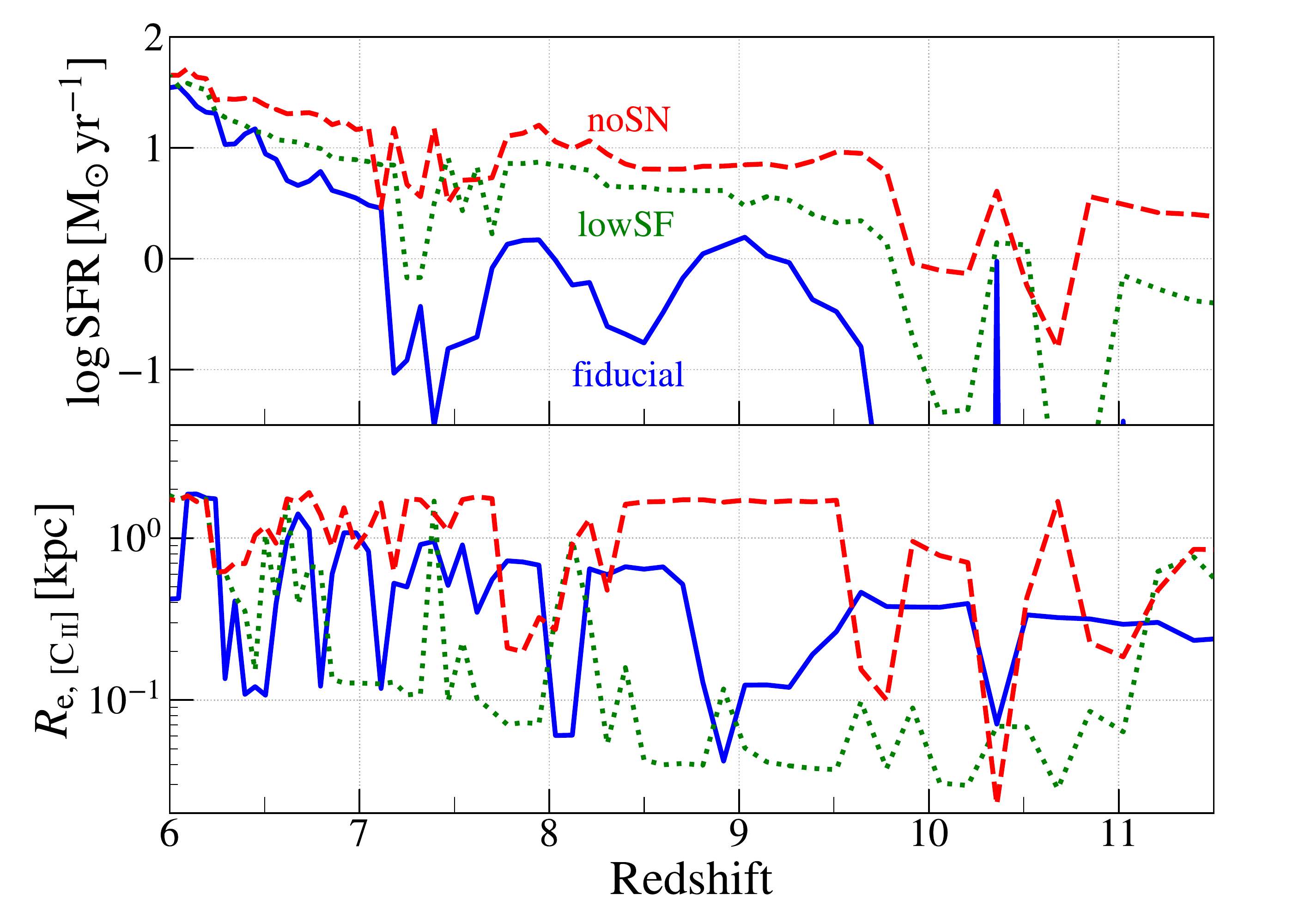}
\caption{Redshift evolution of SFRs ({\it top panel}) and half-light radii of \cii intensity maps ({\it bottom panel}) in the case of Halo-11 (blue solid), Halo-11-noSN (red dashed) and Halo-11-lowSF (green dotted) runs.
We set an upper limit of $R_{\rm e,[C_{\II}]}$ at $0.1r_{\rm vir}$ to avoid artificially large sizes due to galaxy mergers.
}
\label{fig:sizeevo}
\end{center}
\end{figure}

Recent ALMA observations have allowed us to study the kinematics and sizes of distant galaxies via \cii emission.
Thus we here focus on the galaxy size measured by \cii 158\,${\rm \mu m}$ line.
Figure\,\ref{fig:sizeevo} shows that the half-light radius of {\cii} intensity map changes with intermittent star formation.
It is dominated by the emission from central high-density clumps and becomes small ($\sim 0.1\,{\rm kpc}$) in star-burst phases, while it becomes extended in the outflow phase.
In this figure, we set an upper limit to the half-light radius at $0.1r_{\rm vir}$ ($\propto \Mh^{1/3}(1+z)^{-1}$) to avoid artificially large sizes due to galaxy mergers.
At $z<8$, \cii sizes rapidly changes due to complex clumpy gas structures in the disk.

In the lowSF case (green dotted), the inefficient star formation (lower-$A$ coefficient in the Kennicutt--Schdmit relation) induces higher gas density at the galactic center, because the gas is transported before the star formation proceeds. The high-density regions can keep the higher SFR due to lower feedback efficiency via efficient radiative cooling (see also Y17 and A19).
We find that \cii size in the lowSF case has lower values than that of the fiducial case, because high-density clumps continuously reside in the central region (see also, Sec.\,\ref{sec:model}). 

\begin{figure}
\begin{center}
\includegraphics[width=\columnwidth]{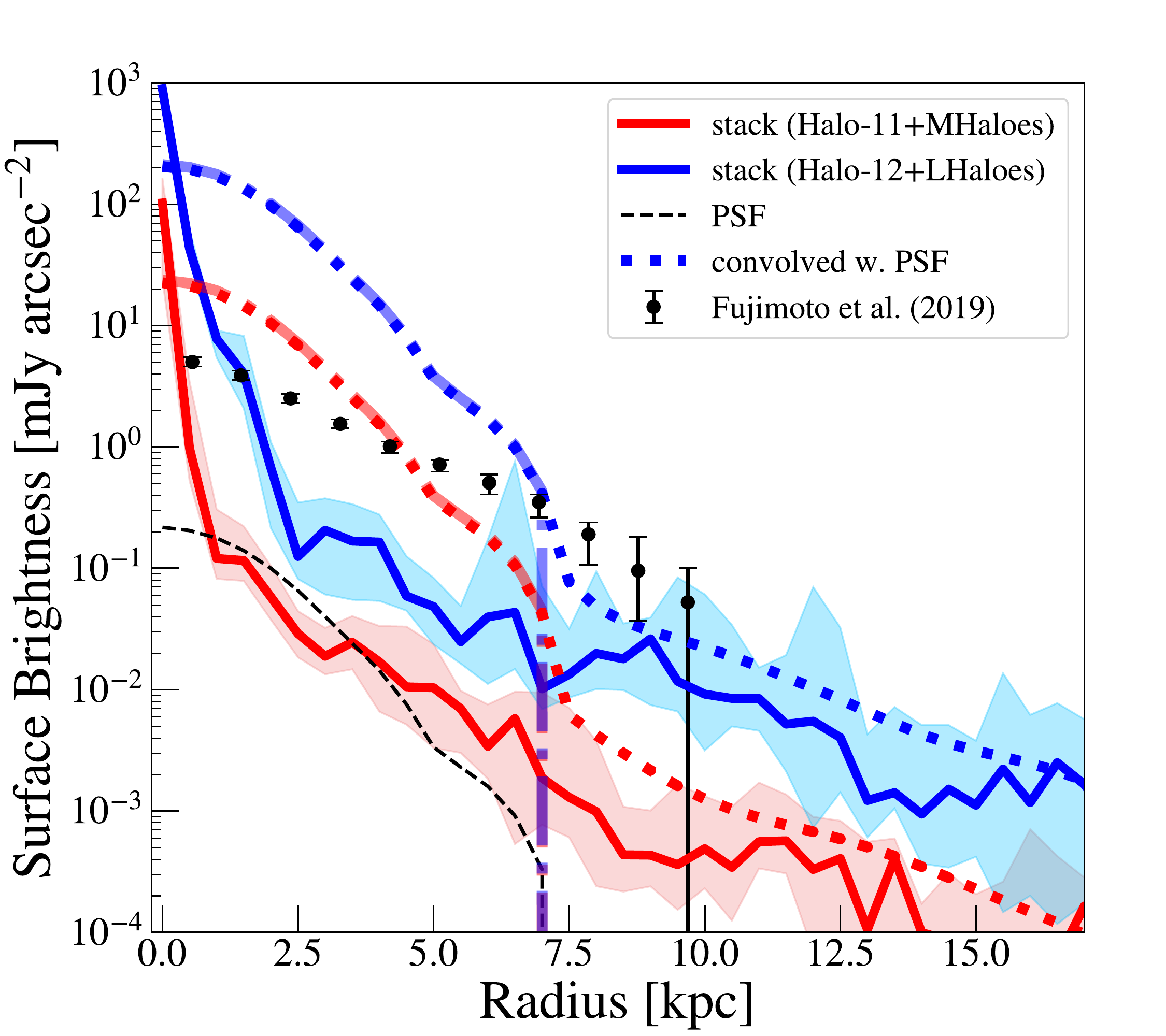}
\caption{Radial profile of \cii surface brightness. The red solid line and the shade represent median and quartiles for 15 stacked samples ($[$Hal-11 and MHaloes$]\,\times$\,3-viewing angles) at $z\sim 6$. The blue solid line and the shade are for the stacked data of $[$Halo-12 and LHaloes$]\,\times$\,3-viewing angles. 
The black points and errorbars represent observational stacking result over 18 galaxies at $z\sim 6$  \citep{Fujimoto19}.
The simulation profiles are very peaky at the center, and become the dotted and dot-dashed lines after convolving with the point spread function (PSF) for ALMA (see text).
}
\label{fig:rp_cii}
\end{center}
\end{figure}

Figure\,\ref{fig:rp_cii} presents the stacked \cii radial profiles for $\Mh\sim 10^{11}\,\Msun$ haloes (red; Halo-11\,+\,MHaloes), and for $\Mh\sim 10^{12}\,\Msun$ haloes (blue; Halo-12\,+\,LHaloes) at $z\sim 6$.
Here we make the \cii intensity maps seen from three viewing angles for each halo, and examine the variation. 
Our simulations show that the surface brightness extension becomes large as the galaxy mass increases.
The two solid lines in Fig.\,\ref{fig:rp_cii} show that the intrinsic intensity profiles are very peaky at the inner radii ($r\lesssim 3\,{\rm kpc}$), and has extend wings in the outskirts which decreases slowly. 
To mimic the observation by ALMA, we convolve the simulated profiles with the point spread function (PSF) of ALMA in two different ways. 
The ALMA PSF has a gaussian-like central peak and a long tail with slight negative values (Fujimoto, in private communication).  
The first method is to fully consider the negative tail of ALMA PSF at $r>7\,{\rm kpc}$, which causes abrupt truncation of the convolved profile at $r\sim 7$\,kpc as shown by the dot-dashed lines.
The second method is to simply neglect the negative part of PSF and only use the positive gaussian-like peak in the central part for convolution, which results in the dotted curve. 
In the second method, the central profile is identical with the first method at $r<7$\,kpc, but the extended outer tail is preserved after the convolution without being subtracted to zero. 
In both cases, the PSF is normalized to unity, so that the total flux is conserved before and after the convolution. 
Both cases show that the shape of inner profiles ($r<7\,{\rm kpc}$) is completely determined by that of the PSF (because the central intrinsic peak is acting almost like a delta-function), and the contribution from outer radii are negligible.
The observational result by \citet{Fujimoto19} is shown by the black solid line, which extends out to $r\sim 9.5$\,kpc with a shallower slope than the simulation results. 
They used galaxies with $-23\lesssim M_{\rm UV}\lesssim -21$, which is in the similar range of simulated galaxies in our $\Mh\sim 10^{12}\,\Msun$ haloes (see Table\,\ref{table:setup}).
The average \cii luminosity of observed galaxies ($\sim 6.3\times 10^{8}\,\Lsun$) is  intermediate between the averages of Halo-11 + MHaloes ($\sim 3.7\times 10^{8}\,\Lsun$) and Halo-12 + LHaloes ($\sim 5.2\times 10^{9}\,\Lsun$).
Even after the convolution, we find that the simulation results are much steeper than the observed result with higher intensities in the central region.  
The intensity profile in the outer part ($r>7$\,kpc) is similar to the observed one by \citet{Fujimoto19}, but is slightly lower at $r\sim 10$\,kpc. 
These differences in the intensity profiles might give us important clues regarding the physics of feedback in high-$z$ galaxies. 
In order for our simulations to reproduce the observation, the \cii flux has to be decreased in the central region by about $1-2$ orders of magnitude.
Using the escape-probability program RADEX \citep{vanderTak07}, \citet{Neri14} suggested that the \cii line of high-$z$ galaxies might be optically thick at a C$^+$ column density of $\gtrsim 10^{18}\,{\rm cm^{-2}}$ \citep[see also,][]{Crawford85,Stacey91a,Stacey91b,Mashian13}, which corresponds to a hydrogen column density $\gtrsim 10^{21.6}\,{\rm cm^{-2}}$ (which further corresponds to the SFR density of $\sim 2.6\times 10^{-2}\,{\rm \Msun\,yr^{-1}\,kpc^{-2}}$). Here we assume the solar abundance ratio \citep[$12+\log{\rm (C/H)}=8.43$,][]{Asplund09}.
If the metallicity is less than  solar, the critical hydrogen column density should be higher, and we can estimate it as $N_{\rm H, crit} \sim 10^{22.6}\left(Z/0.1\,\Zsun\right)^{-1}~{\rm cm^{-2}}$.
The central column densities of our simulated galaxies are $N_{\rm H}\sim 10^{23}\,{\rm cm^{-2}}$ and the metallicities are higher than $0.1\,\Zsun$. Thus we conclude that the assumption of optically-thin emission in our model would overestimate the central \cii luminosity.
In addition, our simulations do not resolve molecular cloud formation in which C$^{+}$ ions will rapidly react to form CO molecules and reduce further central \cii emission \citep{Narayanan17}.

Furthermore, a more realistic feedback model might reduce the discrepancy of \cii luminosity in the outer part.
For star-forming galaxies, stellar radiation pressure ejects the gas out of the disk within a few Myr,
then the cold gas clouds are further accelerated by the ram pressure of SNe.
These processes generate cold outflows with $\sim 10^{4}\,{\rm K}$ \citep[e.g.][]{Murray11, Muratov15}, and the stacking from various viewing angles might produce the extended {\cii} profile.
In addition, we note that there is the difference of gas distribution among the code schemes \citep[e.g.][]{AGORA16}.
In \citet{Fujimoto19}, we compared our simulations with those of \citet{Pallottini17b} which were calculated by an adaptive mesh refinement code {\sc Ramses} \citep{Teyssier02}.
Their galaxies are more extended than ours and matched with observation in the inner part. However, both simulations could not reproduce the extended \cii halo in the outer part. 

Furthermore, using a spherically symmetric outflow model, 
\citet{Pizzati20} studied the \cii profile semi-analytically. They argued that a very high-mass loading factor ($\eta \sim 3.2$) was required to explain the observed profiles. To achieve the high-$\eta$, we might need to introduce additional feedback processes in our simulations.

There are some observational hints from observations on this point already. 
Using the \cii stacked data of $z\sim 5$ galaxies in the ALPINE survey, \citet{Ginolfi19} reported a detection of outflows with a velocity of $\lesssim 500\,{\rm km~s^{-1}}$. They also showed that higher-SFR galaxies have more extended \cii profiles.
Our simulations show that high-SFR galaxies are embedded in massive haloes.
Therefore we expect that more massive haloes have more extended \cii profiles.
We will examine how star formation and feedback models affect \cii profiles of high-$z$ galaxies in our future work, and will compare with above observations further.


%
%
\section{Discussion}
\label{sec:discussion}

\subsection{Impact of Star Formation and SN Feedback Models}
\label{sec:model}

Physical properties of the first galaxies and their UV/IR continuum fluxes can sensitively depend on the models of star formation and SN feedback \citep[e.g.,][]{Wise09,Maio11,Johnson13,Hopkins14,Kimm14,Paardekooper15,Yajima15b,Yajima17,Behrens18,Ma19,Arata19}.
In this subsection, we study the impact of star formation and feedback models on metal emission lines.

\begin{figure}
\begin{center}
\includegraphics[width=\columnwidth]{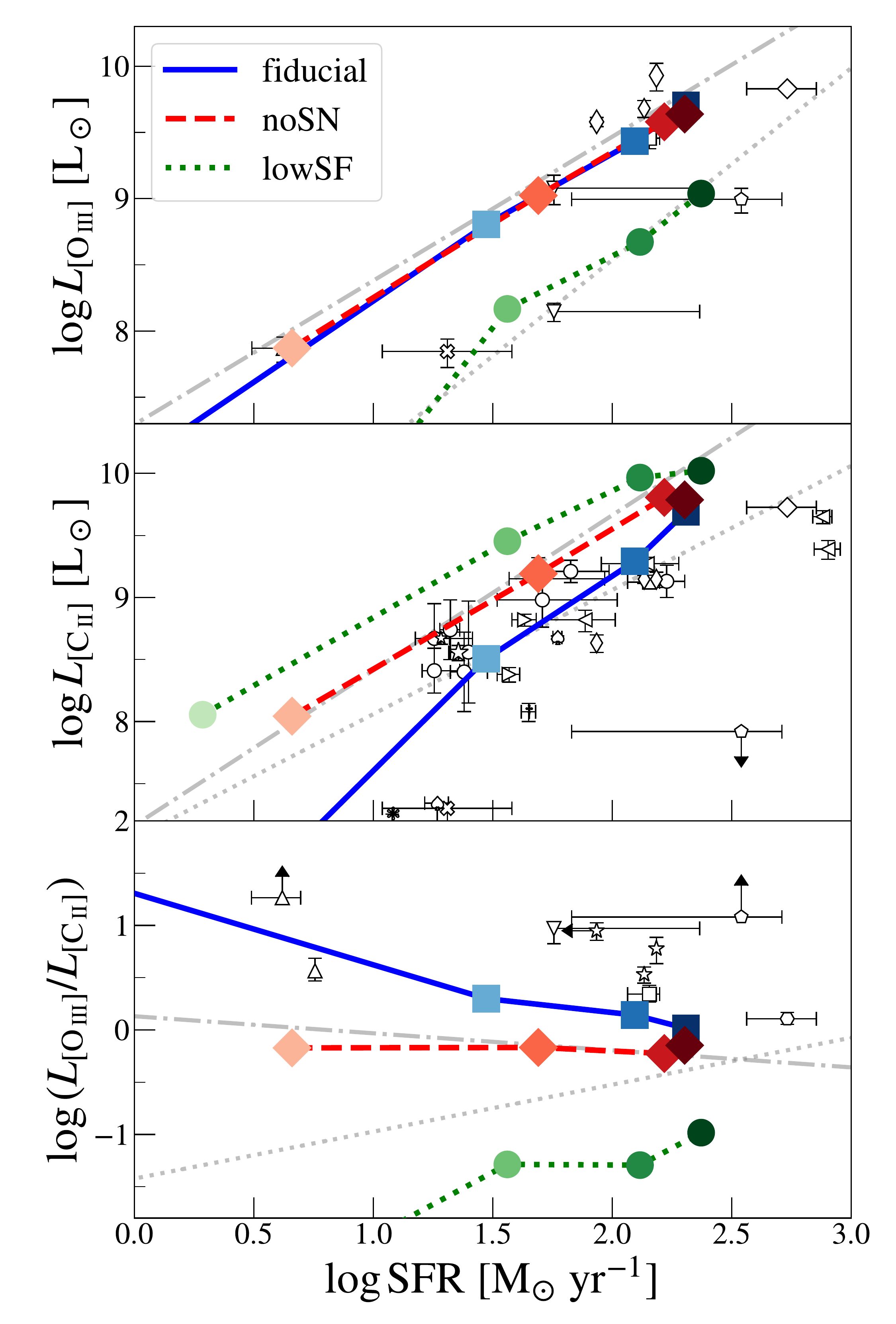}
\caption{Model dependence of relation between SFR and metal line luminosities for Halo-12. 
The blue solid line and squares represent the fiducial case at $z=9,~8,~7$ \& 6 (lighter to darker). 
The red dashed line and diamonds are for the no-SN case, and green dotted line and circles are for the low-SF case. 
The open symbols represent observed galaxies at $z\sim 6-9$ (same as in Fig.\,\ref{fig:oiii} and Fig.\,\ref{fig:cii}).
}
\label{fig:model_lumi}
\end{center}
\end{figure}

Figure\,\ref{fig:model_lumi} compares the \oiii and \cii luminosities from different sub-grid models. 
In the low-SF run, the central gas density becomes higher than in the fiducial case, which results in inefficient SN feedback due to rapid cooling of injected energy \citep{Yajima17}. 
The dense gas also induces rapid hydrogen recombination, resulting in the shrinkage of {\sc H\,ii} regions.  Therefore,  the \oiii luminosity becomes smaller by a factor of $\sim 2$ than in the fiducial case at $z\sim 6$.
Conversely, \cii luminosity in the low-SF case is higher than that of fiducial case due to efficient {\sc H\,i} cloud formation.
Using an analytical model based on the Kennicutt--Schmidt relation, \citet{Ferrara19} showed that a high-$A$ coefficient could explain the lower \cii luminosity of high-$z$ galaxies than local ones, 
because high $\Sigma_{\rm SFR}$ at a specific $\Sigma_{\rm g}$ induces a high ionization degree of ISM. 
Our simulation is consistent with their results.

The central density and SFR in the  no-SN model are similar to the fiducial run during the star-burst phases \citep[see Fig.\,3 in][]{Yajima17}, resulting in similar volume fraction of {\sc H\,ii} regions, and the $L_{\rm [O_{\III}]}$--SFR relations in the two models are very close. 
On the other hand, the \cii luminosity is higher by a factor of few than that of fiducial case, because carbon enrichment has proceeded earlier (see Fig.\,\ref{fig:ratio_halo11}).
Due to these effects the $L_{\rm [O_{\III}]}/L_{\rm [C_{\II}]}$ changes significantly depending on the star formation and feedback models (bottom panel). Only the fiducial case reproduces the observed  negative correlation between the luminosity ratio and SFR.

\begin{figure}
\begin{center}
\includegraphics[width=\columnwidth]{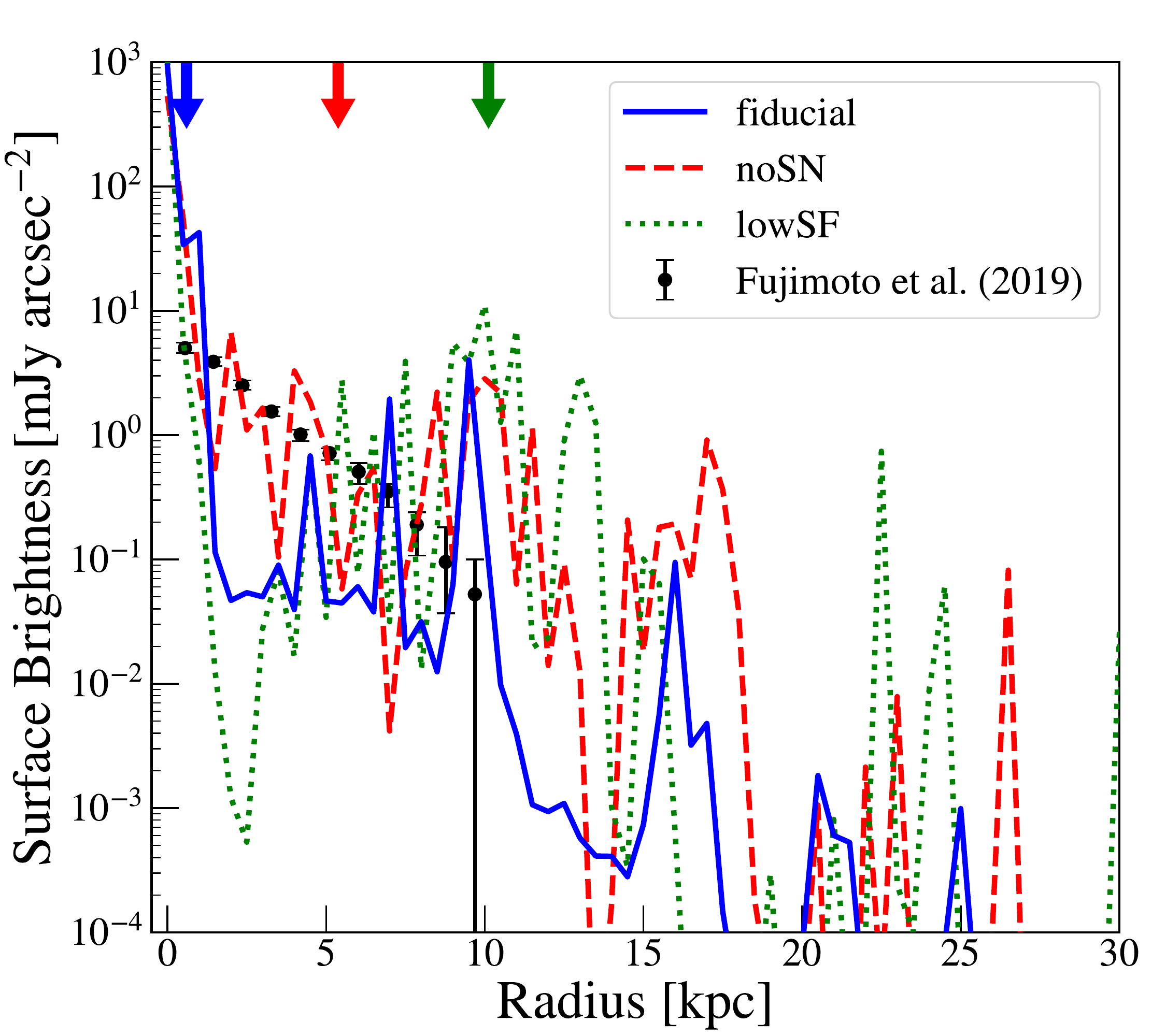}
\caption{Radial profiles of \cii surface brightness of Halo-12 at $z= 6.45$ in fiducial case (blue solid line), no-SN case (red dashed line) and low-SF case (green dotted line). 
The arrows represent the half-light radius in each case.  In the cases of no-SN and low-SF, many {\sc [C\,i\hspace{-.1em}i]}-bright clumps exist in the outer halo, resulting in the extended distribution. 
}
\label{fig:model_radi}
\end{center}
\end{figure}

Figure\,\ref{fig:model_radi} shows the \cii radial profiles for different models at $z=6.45$.
In the fiducial case, the SN feedback destroys {\sc H\,i} clouds in the extended disk, which leads to the concentrated structure.
On the other hand, as described above, the other models can have many {\sc [C\,i\hspace{-.1em}i]}-bright clumps even at the outer halo. 
We find that the half-light radius significantly extends in low-SF and no-SN cases.
Therefore resolving \cii distribution of individual galaxy would be a key to constrain theoretical models.

%
%
\section{Summary}
\label{sec:summary}
Combining zoom-in cosmological hydrodynamic simulations and radiative transfer calculations, we have investigated the radiative properties of galaxies at $z\sim 6-15$ focusing on the \oiii 88\,${\rm \mu m}$ and \cii 158\,${\rm \mu m}$ emission lines.
We use 12 haloes whose masses are $\sim 10^{11}\,\Msun$ (Halo-11 and MHalo-0, 1, 2, 3) and $\sim 10^{12}\,\Msun$ (Halo-12 and LHalo-0, 1, 2, 3, 4, 5) at $z=6$. 
Our major findings are as follows:

\begin{enumerate}
\item
The metal line luminosities rapidly change with intermittent star formation histories due to SN feedback and gas accretion.
The \oiii line is emitted only during the star-burst phases because O$^{2+}$ ions exist in {\sc H\,ii} regions formed by massive stars, while the \cii line is continuously emitted from neutral gas even during the  outflowing phases (Fig.\,\ref{fig:halo11}). 
We show that, in the case of haloes of $\sim 10^{11}~\Msun$,  the \oiii luminosity changes between $\sim 10^{40}-10^{42}\,{\rm erg~s^{-1}}$ at $z<10$.

\vspace{0.2cm}
\item
We find that $\log{L_{\rm [O_{\III}]}}$ is linearly proportional to $\log{\rm SFR}$ (Fig.\,\ref{fig:oiii}).
The relation is very close to that of local metal-poor galaxies \citep{DeLooze14}.
Meanwhile, the $\log{L_{\rm [C_{\II}]}}$--$\log{\rm SFR}$ relation is steeper (slope of $\sim 1.4$) than that of local ones ($\sim 1-1.2$) (Fig.\,\ref{fig:cii}).
We provide fitting formulae for these relations in \S\,\ref{sec:lum_SFR}. 
The \oiii surface brightness is $\gtrsim 1\,{\rm mJy~arcsec^{-2}}$ with very compact and high-density regions at $z=6$. The comparison of size and surface brightness of \oiii regions with observations can help to constrain physical model of feedback from massive stars.

\vspace{0.2cm}
\item
In our samples, line luminosity ratio $L_{\rm [O_{\III}]}/L_{\rm [C_{\II}]}$ decreases  from $\sim 10$ to $\sim 1$ with increasing bolometric luminosity ($10^{9}\,\Lsun - 10^{12}\,\Lsun$) and metallicity ($0.1\,\Zsun - 1\,\Zsun$), which is in good agreement with local and high-$z$ observations (Fig.\,\ref{fig:ratio_all}).
We find that $\log{\rm (O/C)}$ abundance ratio is initially dominated by the oxygen enrichment of Type-II SNe, and decreases from $\sim 0.9$ to $\sim 0.5$ due to carbon-rich wind from AGB stars.
Thus, the $L_{\rm [O_{\III}]}/L_{\rm [C_{\II}]}$ ratio decreases with galaxy evolution and metal enrichment.
We provide fitting formula for the relation between $L_{\rm [O_{\III}]}/L_{\rm [C_{\II}]}$ and metallicity, which could be used to constrain the metallicity of high-$z$ galaxies in the future from the line ratio. 

\vspace{0.2cm}
\item
The luminosity ratio $\log{(L_{\rm [O_{\III}]}/L_{\rm IR})}$ of our samples weakly correlates with $\log{L_{\rm IR}}$ with the slope of $\sim -0.14$, and the negative trend is consistent with high-$z$ observations.
We find that the relation depends on the escape fraction of UV photons.
In the case of low-mass galaxies, most of UV photons can escape without dust absorption. Therefore, the conversion fraction from UV to IR radiation via  dust absorption increases with the galaxy mass.
This results in making the $\log{L_{\rm IR}}$--$\log{\rm SFR}$ relation steeper (slope of $\sim 1.2$).
On the other hand, we see a positive correlation in the $\log{(L_{\rm [C_{\II}]}/L_{\rm IR})}$--$\log{L_{\rm IR}}$ relation, which is inconsistent with observations (Fig.\,\ref{fig:cii_ratio}).
We argue that this inconsistency might be alleviated using a more detailed dust model.
\vspace{0.2cm}
\item
To clarify the impact of sub-grid models of star formation and SN feedback on our results, 
we examine  simulations with lower star formation efficiency (lowSF) and without supernova feedback (noSN).
We find that, in the lowSF case, {\sc H\,ii} regions are not as much extended due to high-density gas and rapid recombination, which results in lower $L_{\rm [O_{\III}]}$ and higher $L_{\rm [C_{\II}]}$ than in the fiducial case by order unity.
In the noSN case, the density structure is similar to that of the fiducial case, resulting in similar \oiii luminosities.
However, the galaxies in the noSN model experience rapid metal enrichment by Type-II SNe and AGB stars and have high carbon abundances. Therefore the \cii luminosity is higher than in the fiducial run.
We find that only the fiducial runs reproduce the observed negative correlation between $L_{\rm [O_{\III}]}/L_{\rm [C_{\II}]}$ and $L_{\rm bol}$.

\end{enumerate}
%
%
%
%
\section*{Acknowledgments}
We thank the referee for the constructive comments. We are grateful to Dr. Inoue for helpful comments on model calculations, and to Dr. Fujimoto for useful discussions on ALMA PSF. 
Numerical computations were carried out on the Cray XC30 \& XC50 at the Center for Computational Astrophysics, National Astronomical Observatory of Japan, and the {\small OCTOPUS} at the Cybermedia Center, Osaka University as part of the HPCI system Research Project (hp180063, hp190050). 
This work is supported in part by the MEXT/JSPS KAKENHI Grant Number JP17H04827, 18H04570, 20H04724 (H.Y.) and JP17H01111, 19H05810 (K.N.), and NAOJ ALMA Scientific Research Grant Number 2019-11A.
KN acknowledges the travel support from the Kavli IPMU, World Premier Research Center Initiative (WPI), where part of this work was conducted. 
Data availability: Data available on request.

%
%
\bibliographystyle{mn}

\bibliography{refs}

%
%
\appendix

\section{Comparison with {\sc Cloudy}}
\label{sec:cloudy}

We compare \oiii 88\,${\rm \mu m}$ luminosity in our model with that of {\sc Cloudy} model \citep[e.g.][]{Ferland98,Inoue11,Inoue14}.
The {\sc Cloudy} table gives us luminosity ratio of \oiii 88\,${\rm \mu m}$ to H$\beta$ lines as a function of density ($n$), ionization parameter ($U\equiv \Phi/n_{\rm e}c$, where $\Phi$ is flux of hydrogen ionizing photons), and metallicity ($Z$), which is based on radiative transfer calculations assuming plane-parallel geometry and pressure equilibrium. 
The range of parameter are $Z/\Zsun= 5\times 10^{-3}-2.5$, $U=10^{-4}-10^{-1}$ and $n=10-10^{3}\,\cc$.
The input stellar spectrum is produced by the {\sc Starburst99} with Salpeter IMF [$0.1-100\,\Msun$], constant star formation for $10\,{\rm Myr}$, and the same stellar metallicity with gaseous component \citep{Inoue14}.
The {\sc Cloudy} table uses the solar elemental abundance ratio.
We use the values of $n,U,Z$ in each cell, which are the result of ionization RT calculation by {\sc Art$^{2}$} code, and refer to the {\sc Cloudy} table for \oiii emissivity ($\epsilon_{\rm cl} \equiv L_{\rm [O_{\III}]}^{\rm cl}/L_{\rm H\beta}^{\rm cl}$).

To obtain \oiii emissivity in our model ($\epsilon$), we divide $L_{\rm [O_{\III}]}$ by H$\beta$ luminosity,  $L_{\rm H\beta}=\alpha_{\rm H\beta}^{\rm eff}(T) n_{\rm p} n_{\rm e} h\nu_{\rm H\beta} V_{\rm cell}$, where $\alpha_{\rm H\beta}^{\rm eff}(T)$ is the effective recombination coefficient \citep{Brocklehurst71} and $n_{\rm p}$ ($n_{\rm e}$) is the proton (electron) density in the cell.
We note that the ${\rm H\beta}$ attenuation by dust grains does not affect significantly because the dust column density in the central regions of our galaxies is $\sim 10^{-3}\,(Z/\Zsun)~{\rm g\,cm^{-2}}$ corresponding to the optically-thin regime for the ${\rm H\beta}$ line.
Although the {\sc Cloudy} table takes the attenuation effect into account, it does not  contribute to $\epsilon_{\rm cl}$ significantly in the range of our input parameters.

\begin{figure}
\begin{center}
\includegraphics[width=\columnwidth]{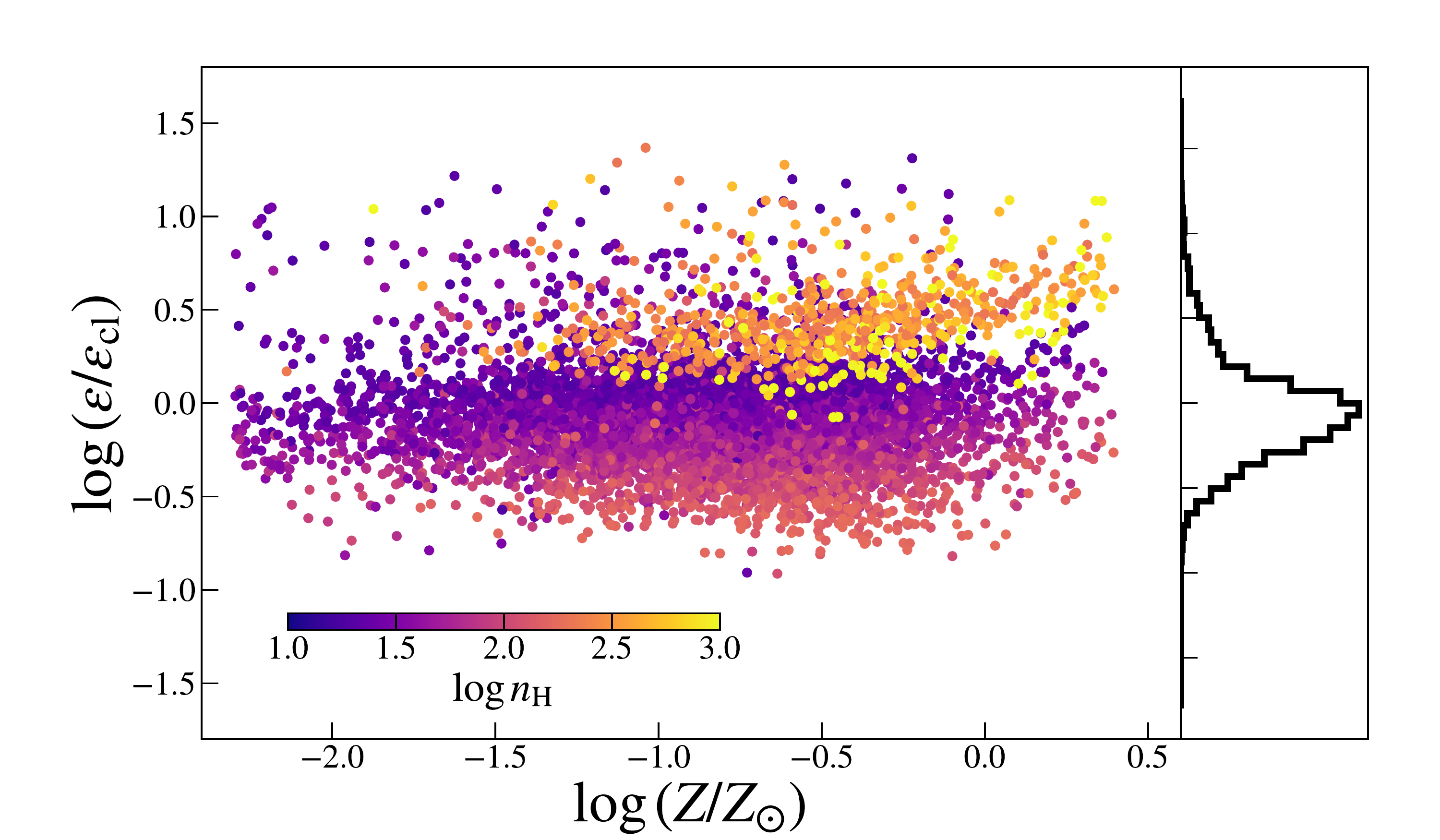}
\caption{Comparison of \oiii emissivity in our model and {\sc Cloudy}. Left panel shows the distribution of emissivity ratio $\epsilon/\epsilon_{\rm cl}$, where $\epsilon$ and $\epsilon_{\rm cl}$ is the luminosity ratio of \oiii 88$\,{\rm \mu m}$ and ${\rm H\beta}$ lines computed by each model, in Halo-11 at $z=6.0$ as a function of metallicity. The color is scaled by gas density. Right panel shows probability distribution function of $\log{(\epsilon/\epsilon_{\rm cl})}$.
}
\label{fig:cloudy}
\end{center}
\end{figure}

Figure\,\ref{fig:cloudy} shows the ratio of $\epsilon$ to $\epsilon_{\rm cl}$ as a function of metallicity.
For most of the cells, $\epsilon/\epsilon_{\rm cl}$ is almost unity, which implies that our simple emission model is in good agreement with {\sc Cloudy}. Total \oiii luminosity is dominated by the contribution from cells with $|\log{\epsilon/\epsilon_{\rm cl}}| < 0.5$ ($\sim 72\,\%$).
Here we focus on the origin of the differences between our model calculation and {\sc Cloudy}.
There are two-types of outliers at  $\log(\epsilon/\epsilon_{\rm cl}) > 0.5$: low-$U$ cells ($\log{U}\sim -4$) and high-density cells ($n \gtrsim 500\,\cc$).
In the former case, {\sc Cloudy} predicts that weak radiation field ionizes the surface of plane-parallel gas ($N \lesssim 10^{19}\,{\rm cm^{-2}}$), and \oiii is emitted only from the thin layer.
Meanwhile, if the electron fraction of a cell is higher than $0.5$, our model calculates optically-thin oxygen ionization equilibrium in the whole cell volume and obtains O$^{2+}$ abundance, which would overestimate \oiii emissivity.
In the later case (indicated by yellow points in Fig.\,\ref{fig:cloudy}), the emissivity in {\sc Cloudy} rapidly decreases with increasing density, because the density exceeds the critical density for transition of ${\rm ^{3}P_{1}}$ $\to$ ${\rm ^{3}P_{0}}$.
Meanwhile, our model predicts that the critical density is higher by a factor of $\sim 3$ ($1.74\times 10^{3}\,\cc$ at $T=10^{4}\,{\rm K}$), thus the emissivity increases with density.
The difference probably comes from referred collision strength $\Omega$ \citep{Aggarwal99}.

As described above, {\sc Cloudy} uses input spectrum assuming a simple stellar population.
However, the actual SED would be the sum of various stellar populations with different ages and metallicities, which  could be more complex. 
Our model calculates O$^{2+}$ abundance under the complex  SEDs, and the oxygen enrichment by SNe and AGB stars is tracked by particles separately.
Thus, we argue that, at least for the SED treatment, our model is doing a more appropriate treatment for estimating \oiii emission in high-$z$ galaxies. 

\section{The CMB effect}
\label{sec:cmb}

Here we estimate the CMB effect onto \cii $158\,{\rm \mu m}$ luminosity \citep[e.g.][]{Goldsmith12}. The population of two energy levels is determined by the rate equation:
\begin{equation}
\label{eq:rate}
    n_{\rm u}(A_{\rm ul}+ B_{\rm ul}J_{\rm \nu}+C_{\rm ul}) 
    = n_{\rm l}(B_{\rm lu}J_{\rm \nu}+C_{\rm lu}),
\end{equation}
where $n_{\rm u} (n_{\rm l})$ is number density of C$^{+}$ ions in the upper (lower) level, and $J_{\rm \nu}$ is the mean intensity of background radiation at 158\,${\rm \mu m}$. 
We assume an escape probability of $\beta=1$, and a  black-body spectra $J_{\rm \nu}=B_{\rm \nu}(T_{\rm CMB})$. The $C_{\rm lu}$ term is sum of collision rate with electrons ($e^{-}$), hydrogen atoms (${\rm H^{0}}$) and hydrogen molecules (${\rm H_{2}}$) \citep{Glover07}. 
The $C_{\rm lu}$ and $C_{\rm ul}$ are related to each other by detailed balance.  The emergent luminosity  is calculated by Equation (\ref{eq:lumi}). 
In diffuse ISM, the stimulated absorption rate ($n_{\rm l}B_{\rm lu}J_{\rm \nu}$) is higher than stimulated emission rate ($n_{\rm u}B_{\rm ul}J_{\rm \nu}$), thus the CMB increases $n_{\rm u}/n_{\rm l}$ ratio and reduces \cii luminosity.

\begin{figure}
\begin{center}
\includegraphics[width=\columnwidth]{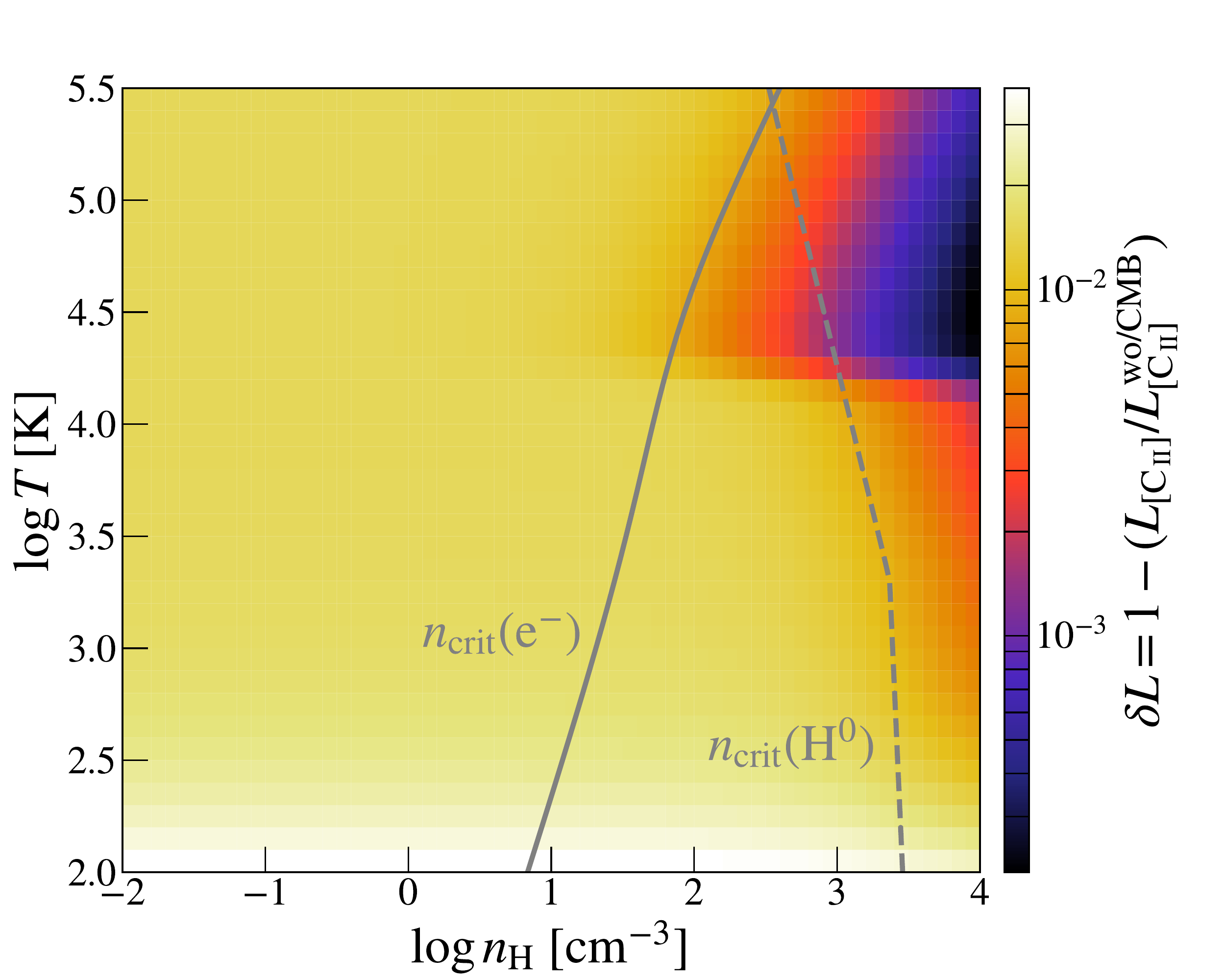}
\caption{The CMB effect to \cii 158\,${\rm \mu m}$ luminosity at $z = 7.0$ as a function of hydrogen nuclei density and temperature. The color scale indicates the reduction strength, $\delta L = 1-(L_{\rm [C_{\,\II}]}/L_{\rm [C_{\,\II}]}^{\rm wo/CMB})$, where $L_{\rm [C_{\,\II}]}$ ($L_{\rm [C_{\,\II}]}^{\rm wo/CMB}$) is the emergent \cii luminosity with (without) the CMB stimulated emission and absorption. Gray solid line and dashed line represent the critical densities of electrons ($n_{\rm crit}({\rm e^{-}})$) and hydrogen atoms ($n_{\rm crit}({\rm H^{0}})$), respectively.
}
\label{fig:cmb}
\end{center}
\end{figure}

Figure~\ref{fig:cmb} shows how much \cii emission is attenuated by CMB at $z= 7.0$. 
For a low temperature gas ($T\lesssim 100\,{\rm K}$), the CMB effect becomes significant compared to collisional excitation, resulting in a reduction of luminosity  $\gtrsim 3\,\%$ \citep[see also,][]{Lagache18}.
At $T\sim 10^{4}\,{\rm K}$, the densities of $e^{-}$ and ${\rm H^{0}}$ dramatically change due to hydrogen recombination, thus the collision partner for C$^{+}$ ions switches. If gas density is higher than the critical density of the collision partner, the rate equation is dominated by the collision terms, and as a result, the CMB effect becomes negligible.
Note that gas temperature in our simulations is higher than $10^{3}\,{\rm K}$ (Fig.\,\ref{fig:phase}), thus the CMB affects \cii luminosity only by a few percent.

Additionally, note that the CMB at $z=0$ also reduces the detectability due to obscuring the line signal by the background noise \citep{daChunha13,Kohandel19}.
We do not consider this effect, because we focus on the intrinsic luminosity of high-$z$ galaxies. However, it could be important when comparing the line profiles in detail between simulations and observations. We will take this effect into account in a future paper.

\label{lastpage}

\end{document}